\documentclass[12pt]{article}
\usepackage{euscript,amsmath, amssymb, amsfonts}
\usepackage{color}

\newcommand{ \tto}{{\longrightarrow}}

\newcommand{\R}{{\mathbb R}}
\newcommand{ \nn}{\nonumber \\}
\newcommand{ \bee}{\begin{eqnarray}}
\newcommand{ \eee}{\end{eqnarray}}

\newcommand{\p}{{\hbar^2}}

\newcommand{\oC}{{\mathbb C}}

\newcommand{\oP}{{\cal P}}
\newcommand{\di}{{ d }}
\newcommand{\D}{\EuScript D}
\newcommand{\eE}{\EuScript E}

\newcommand{\G}{{\mathbb G}}
\newcommand{\Z}{{\mathbb Z}}
\newcommand{\K}{{\mathbb K}}
\newcommand{\be}{\begin{equation}}
\newcommand{\ee}{\end{equation}}

\newcounter{theorem}
\newcommand{\theorem}{\par\refstepcounter{theorem}
           {\bf Theorem \arabic{section}.\arabic{theorem}. }}
\renewcommand\thetheorem{\thesection.\arabic{theorem}}
\makeatletter \@addtoreset{theorem}{section}

\newcounter{lemma}
\newcommand{\lemma}{\par\refstepcounter{lemma}
           {\bf Lemma \arabic{section}.\arabic{lemma}. }}
\renewcommand\thelemma{\thesection.\arabic{lemma}}
\makeatletter \@addtoreset{lemma}{section}

\newcounter{proposition}
\newcommand{\proposition}{\par\refstepcounter{proposition}
           {\bf Proposition \arabic{section}.\arabic{proposition}. }}
\renewcommand\theproposition{\thesection.\arabic{proposition}}
\makeatletter \@addtoreset{proposition}{section}

\makeatletter \@addtoreset{equation}{section}
\def\theequation{\thesection.\arabic{equation}}
\makeatother

\newcounter{appen}
\newcommand{\appen}[1]{\par\refstepcounter{appen}
{\par\medskip\noindent\Large\bf Appendix \arabic{appen}.
\medskip
}{\Large\bf #1}}
\newenvironment{proof}[1][Proof]{\noindent\textsf{#1.\ }}{ \ \rule{0.5em}{0.5em}}

\newcounter{subappen}

\makeatletter \@addtoreset{subappen}{appen}

\newcounter{subsubappen}

\makeatletter \@addtoreset{subsubappen}{subappen}

\font\frtnfr=eufm10   scaled\magstep1
\font\twlfr=eufm10
\font\tenfr=eufm10
\font\egtfr=eufm8
\font\sixfr=eufm6
\newfam\frfam
\textfont\frfam=\frtnfr
\scriptfont\frfam=\twlfr
\scriptscriptfont\frfam=\tenfr
\def\fr{\fam\frfam}

\font\frtnopen=msbm10  scaled\magstep2
\font\twlopen=msbm10
\font\tenopen=msbm10
\font\egtopen=msbm8
\font\sixopen=msbm6
\newfam\openfam
\textfont\openfam=\frtnopen
\scriptfont\openfam=\twlopen
\scriptscriptfont\openfam=\tenopen
\def\open{\fam\openfam}

\font\frtnsf = cmss12 scaled\magstep1
\font\twlsf = cmss10
\font\tensf = cmss9
\font\egtsf = cmss8
\font\sixsf = cmss8
\newfam\Scfam
\textfont\Scfam = \frtnsf
\scriptfont\Scfam = \twlsf
\scriptscriptfont\Scfam = \tensf
\def\Sc{\fam\Scfam}

\begin{document}

\sloppy \title
 {
     Cohomology of antiPoisson superalgebra
 }
\author
 {
 S.E.Konstein\thanks{E-mail: konstein@lpi.ru}\ \ and
 I.V.Tyutin\thanks{E-mail: tyutin@lpi.ru}
 \thanks{
               This work was supported
               by the RFBR (grants No.~05-01-00996
               (I.T.) and No.~05-02-17217 (S.K.)),
               and by the grant LSS-1578.2003.2.
 } \\
               {\sf \small I.E.Tamm Department of
               Theoretical Physics,} \\ {\sf \small P. N. Lebedev Physical
               Institute,} \\ {\sf \small 119991, Leninsky Prospect 53,
               Moscow, Russia.} }
\date {}

\maketitle

\begin{abstract}
{ \footnotesize We consider antiPoisson superalgebras realized on the smooth
Grassmann-valued functions with compact support in $\R^n$ and with the grading inverse
to Grassmanian parity. The lower cohomologies of these superalgebras
are found.
}
\end{abstract}


\section{Introduction}

The odd Poisson bracket play an important role in Lagrangian
formulation of the quantum theory of the gauge fields, which is
known as BV-formalism \cite{BV1}, \cite{BV2}
(see also \cite{reports}-\cite{books}).
In \cite{L}, it was shown that there are two  analogs of the Poisson bracket
and related \lq\lq mechanics": a direct one, still called the Poisson bracket,
and the ``odd'' one,
introduced in physical literature in \cite{BV1}
under the name ``antibracket".

The antibracket possesses many features analogous to
those of the Poisson bracket and even can be obtained via \lq\lq
canonical formalism" with an \lq\lq odd time". However, unlike the
Poisson bracket, on different aspects of whose deformations
(quantization) there is voluminous literature, the deformations of
the antibracket is not satisfactorily studied yet. The only result
is \cite{Leites}, where the deformations of the Poisson and
antibracket superalgebras realized on the superspace of {\it
polynomials} are found.

The goal of present work is finding the lower cohomology spaces
of antiPoisson superalgebra realized on
the {\it smooth} Grassmann-valued functions with
compact support in $\R^n$. These results is used in the next work
\cite{next} where the general form of the deformation of such
antiPoisson superalgebra is found. Particularly, it is shown in
\cite{next} that the nontrivial deformations do exist.

Let $\K$ be either $\R$ or $\oC$. We denote by ${\cal D}(\R^n)$ the space
of smooth $\K$-valued functions with compact support on $\R^n$. This space
is endowed with its standard topology. We set
$$
\mathbf D^{n_-}_{n_+}= {\cal
D}(\R^{n_+})\otimes \G^{n_-},\quad \mathbf E^{n_-}_{n_+}=
C^\infty(\R^{n_+})\otimes \G^{n_-},\quad \mathbf D^{\prime n_-}_{n_+}=
{\cal D}'(\R^{n_+})\otimes \G^{n_-},
$$
where $\G^{n_-}$ is the Grassmann algebra with $n_-$ generators and
$\EuScript D'(\R^{n_+})$ is the space of continuous linear
functionals on $\EuScript D(\R^{n_+})$. The generators of the
Grassmann algebra (resp., the coordinates of the space $\R^{n_+}$)
are denoted by $\xi^\alpha$ for $\alpha=1,\ldots,n_-$ (resp., $x^i$
for $i=1,\ldots, n_+$). We shall also use common notation $z^A$
which are equal to $x^A$ for $A=1,\ldots,n_+$ and to $\xi^{A-n_+}$
for $A=n_++1,\ldots,n_++n_-$.

The spaces $\mathbf D^{n_-}_{n_+}$, $\mathbf E^{n_-}_{n_+}$, and
$\mathbf D^{\prime n_-}_{n_+}$ possess a natural parity which is
determined by that of the Grassmann algebra, it is denoted by
$\varepsilon$. Set: $\epsilon=\varepsilon+1$.

We set $\varepsilon_A=0$ for $A=1,\ldots, n_+$ and $\varepsilon_A=1$
for $A=n_++1,\ldots, n_++n_-$.

It is well known, that if $n_+=n_-=n$ then the bracket
\bee\label{Sch}
[f,g](z)=\sum_{i=1}^n\left(f(z)\frac{\overleftarrow{\partial}}{\partial
x^i} \frac\partial{\partial\xi^i}g(z)-
f(z)\frac{\overleftarrow{\partial}}{\partial\xi^i}
\frac{\partial}{\partial x^i}g(z)\right), \eee called {\it
antibracket}, defines a Lie superalgebra structure on the
superspaces $\mathbf D_{n}\stackrel {def} = \mathbf D^{n}_{n}$ and
$\mathbf E_{n}\stackrel {def} = \mathbf E^{n}_{n}$ with the
$\epsilon$-parity. Clearly, the form $\omega$ defining the
antibracket
$$
[f,g](z)=f(z)\frac{\overleftarrow{\partial}}{\partial z^A}
\omega^{AB}
\frac\partial{\partial z^B}g(z),
$$
is constant, non-degenerate, and satisfies the condition
$$
\omega^{BA}=-(-1)^{\epsilon_A\epsilon_B}\omega^{AB},\;
\epsilon(\omega^{AB})=\epsilon_A+\epsilon_B,
$$

Here these Lie superalgebras are called the {\it antiPoisson superalgebras}.
\footnote{We will also consider the usual
multiplication of the elements of the antiPoisson superalgebras with
commutation relations $fg=(-1)^{\varepsilon(f)\varepsilon(g)}gf$, so
the $x^i$ will be called even variables and the $\xi^i$ will be
the odd ones.} We set: $\mathbf D^\prime_{n} \stackrel {def} =
\mathbf D^{\prime n}_{n}$.

The integral on $\mathbf D_{n}$ is defined by the relation $$\int
\di z\, f(z)= \int_{\R^{n}}\di x\int \di\xi\, f(z), $$ where the
integral on the Grassmann algebra is normalized by the condition
$$\int \di\xi\, \xi^1\ldots\xi^{n}=1.$$ We identify $\G^{n}$ with its
dual space $\G^{\prime n}$ setting $$f(g)=\int\di\xi\,
f(\xi)g(\xi)\text{ for any $f,g\in \G^{n}$}.$$ Accordingly, the
space $\mathbf D^{\prime}_{n}$ of continuous linear functionals on
$\mathbf D_{n}$ is identified with $\D^\prime(\R^{n})\otimes
\G^{n}$. The value $m(f)$ of a functional $m\in\mathbf D^\prime_n$
on a test function $f\in\mathbf D_n$ will be often written in the
integral form:
$$m(f)=\int\di z\,m(z)f(z).$$

\section{Cohomology of the antiPoisson superalgebras (Results)}
Let $\mathbf D_{n}$ act in a $\Z_2$-graded space $V$ (the action of
$f\in \mathbf D_{n}$ on $v\in V$ will be denoted by $f\cdot v$). The
space $C^p(\mathbf D_{n},V)$ of $p$-cochains consists, as in
\cite{SKT1}, of all separately continuous multilinear
superantisymmetric maps $\mathbf D_{n}^p\tto V$. Superantisymmetry
means, as usual, that
$$M_p(\,...\,,f_i,\,f_{i+1},...)=-
(-1)^{\epsilon(f_i)\epsilon(f_{i+1})}M_p(\,...,f_{i+1},f_{i},...).$$
The space $C^p(\mathbf D_{n}, V)$ possesses a natural
$\Z_2$-grading:
$$
\epsilon(M_p(f_1,\ldots,f_p))=
\epsilon_{M_p}+\epsilon(f_1)+\ldots+\epsilon(f_p)
$$
for any (homogeneous) $f_j\in\mathbf D_{n}$. We will often use the
Grassmann $\varepsilon$-parity\footnote{If $V$ is the space of
Grassmann-valued functions on $\R^n$, then $\varepsilon$ defined in
such a way coincides with the usual Grassmann parity. } of cochains:
$\varepsilon_{M_p}=\epsilon_{M_p}+p+1$. The differential $\di_p^V:
C^p(\mathbf D_{n}, V)\tto C^{p+1}(\mathbf D_{n}, V)$ is defined as
follows:
\begin{eqnarray}
&&d_p^{V}M_p(f_1,...,f_{p+1})=
-\sum_{j=1}^{p+1}(-1)^{j+\epsilon(f_j)|\epsilon(f)|_{1,j-1}+
\epsilon(f_j)\epsilon_{M_p}}f_j\cdot
M_p(f_{1},...,\breve{f}_j,...,f_{p+1})- \nonumber \\
&&-\sum_{i<j}(-1)^{j+\epsilon(f_j)|\epsilon(f)|_{i+1,j-1}}
M_p(f_1,...f_{i-1},[f_i,f_j],f_{i+1},...,
\breve{f}_j,...,f_{p+1}),\label{diff}
\end{eqnarray}
for any $M_p\in C^p(\mathbf D_{n}, V)$ and
$f_1,\ldots,f_{p+1}\in\mathbf D_{n}$ having definite
$\epsilon$-parities. Here the sign $\breve{}$ means that the
argument is omitted and
$$
|\epsilon(f)|_{i,j}=\sum_{l=i}^j\epsilon(f_l).
$$
We have $\di^V_{p+1}\di^V_p=0$ for any $p=0,1,\ldots$. The $p$-th
cohomology space of the differential $\di_p^V$ will be denoted by
$H^p_V$. The second cohomology space $H^2_{\mathrm{ad}}$ in the
adjoint representation is closely related to computing infinitesimal
deformations of the Lie bracket $[\cdot,\cdot]$ in the form $$
[f,g]_*=[f,g]+\hbar[f,g]_1+\ldots$$ up to similarity transformations
$$[f,g]_T=T^{-1}[Tf,Tg], $$ where a continuous linear operator $T:V[[\p]]\tto V[[\p]]$
is of the form $T=\mathsf {id}+\p T_1$. The condition that
$[\cdot,\cdot]_1$ is a 2-cocycle is equivalent to the Jacobi
identity for $[\cdot,\cdot]_*$ modulo the $\hbar^2$-order terms.

We study the cohomology of the antiPoisson superalgebra $\mathbf
D_{n}$ in the following cases:
\begin{enumerate}
\item\
The trivial representation: $V=\K$, $f\cdot a=0$ for any $f\in
\mathbf D_{n}$ and $a\in\K$. (Notation: $H^p_{\mathrm{tr}}$ and
$\di^{\mathrm{tr}}_p$).
\item
$V=\mathbf D_{n}'$ and $f\cdot g=[f,g]$ for any $f\in\mathbf D_{n}$,
$g\in\mathbf D_{n}'$. (Notation:  $H^p_{\mathrm{D'}}$ and
$\di^{\mathrm{ad}}_p$).
\item
$V=\mathbf E_{n}$ and $f\cdot g=[f,g]$ for any $f\in\mathbf D_{n}$,
$g\in\mathbf E_{n}$. (Notation: $H^p_{\mathrm{E}}$ and
$\di^{\mathrm{ad}}_p$).
\item
The adjoint representation: $V=\mathbf D_{n}$ and $f\cdot g=[f,g]$
for any $f,g\in\mathbf D_{n}$. (Notation: $H^p_{\mathrm{ad}}$ and
$\di^{\mathrm{ad}}_p$).
\end{enumerate}
In the case of the trivial representation, $\K$ is considered as a
superspace whose $\epsilon$-even subspace is zero. We say that the
$p$-cocycles $M_p^1,\ldots M_p^k$ are {\it independent} if they give
rise to linearly independent elements in $H^p$. For a multilinear
form $M_p$ taking values in $\mathbf D_{n}$, $\mathbf E_{n}$, or
$\mathbf D^{\prime}_{n}$, we write $M_p(z|f_1,\ldots,f_p)$ instead
of more cumbersome $M_p(f_1,\ldots,f_p)(z)$.

The following theorems describe lower cohomology of the antiPoisson superalgebra.
\theorem\label{th1}{} {\it \begin{enumerate}
\item
$H^1_{\mathrm{tr}}\simeq 0$.
\item
Let $n\geq 2$. Then $H^2_{\mathrm{tr}}\simeq 0$.

Let $n=1$. Then $H^2_{\mathrm{tr}}\simeq \K^2$, and the cochains
\be
\mu_1(f,g)=\int dz(-1)^{\varepsilon(f)}\{\partial_x^3\partial_\xi f(z)\}g(z),
\;\; \mu_2(f,g)=\int dz(-1)^{\varepsilon(f)}\{\partial_x^2f(z)\}g(z)
\label{**mu}
\ee
are independent nontrivial cocycles.
\end{enumerate}
}

It follows from Theorem \ref{th1} that if $n=1$, then the antiPoisson
superalgebra has a 2-parametric central extension. These extensions
are described in \cite{next}.

Now, let $\mathbf{Z}_{n}=\mathbf{D}_{n} \oplus {\cal
C}_{\mathbf{E}_{n}}(\mathbf{D}_{n})$, where ${\cal
C}_{\mathbf{E}_{n}}(\mathbf{D}_{n})$ is a centralizer of
$\mathbf{D}_{n}$ in $\mathbf{E}_{n}$. Clearly, ${\cal
C}_{\mathbf{E}_{n}}(\mathbf{D}_{n})=\K$.

\theorem\label{th2}{} {\it
\begin{enumerate}

\item
$H^0_{\mathbf D'}\simeq
H^0_{\mathbf E}\simeq \K$; the function $m_0(z)\equiv 1$ is a nontrivial
cocycle.

$H^0_{\mathrm {ad}}\simeq 0$.

\item\begin{enumerate}\item
$H^1_{\mathbf D'}\simeq
H^1_{\mathbf E}\simeq \K^2$; independent nontrivial cocycles are given by
\[
m_{1|1}(z|f)=\eE_z f(z),\;\; m_{1|2}(z|f)=\Delta f(z),
\]
where
\be
\eE_z = 1-\frac 1 2 z^A \frac \partial {\partial z^A},\;\;
\Delta=\frac 1 2 (-1)^{\varepsilon_A}
\omega^{AB}\frac{\partial}{\partial z^A}\frac{\partial}{\partial z^B}
=\frac \partial {\partial x^i}\frac \partial {\partial \xi^i}.
\label{Delta}
\ee

\item
Let $V_2$ be the two-dimensional subspace of $C^1(\mathbf
D_{n},\mathbf D_{n})$ generated by the cocycles $m_{1|1}$ and
$m_{1|2}$. Then there is a natural isomorphism $V_2\oplus (\mathbf
E_{n}/\mathbf Z_{n})\simeq H^1_{\mathrm{ad}}$ taking $(M_1,T)\in
V_2\oplus (\mathbf E_{n}/\mathbf Z_{n})$ to the cohomology class
determined by the cocycle $M_1(z|f)+[t(z),f(z)]$, where $t\in
\mathbf E_{n}$ belongs to the equivalence class $T$.
\end{enumerate}
\item
Let the bilinear maps $m_{2|1}, \;m_{2|2}:(\mathbf D_1)^2\tto
\mathbf E_{1}$ and $m_{2|3}\;m_{2|4}:(\mathbf D_n)^2\tto\mathbf
E_{n}$ be defined by the relations
\begin{eqnarray}
&&m_{2|1}(z|f,g)=\int du \partial_\eta g(u)\partial^{3}_y f(u),\;\;
\epsilon_{m_{2|1}}=1, \label{5.2.7} \\
&&m_{2|2}(z|f,g)=\int du\theta(x-y)[\partial_\eta g(u)\partial_y^3f(u)-
\partial_\eta f(u)\partial_y^3 g(u)]+ \nonumber \\
&&+x[\{\partial_\xi\partial_{x}^{2}f(z)\}\partial_\xi\partial_{x}g(z)-
\{\partial_\xi\partial_xf(z)\}\partial_\xi
\partial_x^2g(z)],\;\; \epsilon_{m_{2|2}}=1,
\label{5.2.8} \\
&&m_{2|3}(z|f,g)=(-1)^{\varepsilon(f)}\{(1-N_{\xi})f(z)\}(1-N_{\xi})g(z), \;\;
\epsilon_{m_{2|3}}=1, \label{6.3a} \\
&&m_{2|4}(z|f,g)=(-1)^{\varepsilon (f)}\{\Delta f(z)\}\eE_z g(z)+
\{\eE_z f(z)\}\Delta g(z)\;\;  \epsilon_{m_{2|4}}=0. \label{6.3b}
\end{eqnarray}
where $z=(x,\xi)$, $u=(y,\eta)$, $N_\xi=\xi \partial_{\xi}$.

Then

\begin{enumerate}

\item
$H^2_{\mathrm {ad}}\simeq \K^2$ and the cochains
$m_{2|3}(z|f,g)$ and $m_{2|4}(z|f,g)$
are
independent nontrivial cocycles.

\item
Let $n=1$.

Then
$H^2_{\mathbf D'}\simeq
H^2_{\mathbf E}\simeq  \K^4$ and the cochains
$m_{2|1}(z|f,g)$, $m_{2|2}(z|f,g)$, $m_{2|3}(z|f,g)$, and $m_{2|4}(z|f,g)$
are
independent nontrivial cocycles.

\item
Let $n\geq 2$. Then $H^2_{\mathbf D'}\simeq H^2_{\mathbf E}\simeq \K^2$ and
the cochains $m_{2|3}(z|f,g)$ and $m_{2|4}(z|f,g)$ are independent nontrivial
cocycles.

\end{enumerate}
 \end{enumerate}
}

\section{Preliminary and Notation}

We define $\delta$-function by the formula
\[
\int dz^{\prime }\delta (z^{\prime }-z)f(z^{\prime })=\int f(z^\prime)
\delta (z-z^{\prime })dz^{\prime }=f(z).
\]

Evidently,
\[
[f,g](z)=(-1)^{\varepsilon_A\epsilon(f)}\frac{\partial}{\partial z^A}
(f(z)\omega^{AB}\frac{\partial}{\partial z^B}g(z))-2f\Delta g(z),\]
\[
(-1)^{\epsilon(g)}\int dzf[g,h]=\int dz[f,g]h+2\int dzf\Delta gh,
\]
where $\Delta$ is defined by (\ref{Delta}).

The following notation is used below:
\begin{eqnarray}
&& T_{\ldots (A)_{k}\ldots }\equiv T_{\ldots A_{1}\ldots A_{k}\ldots },\quad
T_{\ldots A_{i}A_{i+1}\ldots }=(-1)^{\varepsilon _{A_{i}}\varepsilon
_{A_{i+1}}}T_{\ldots A_{i+1}A_{i}\ldots },\quad i=1,\ldots ,k-1 \nn
&& T_{\ldots (A)_{k}\ldots }Q_{\ldots }{}^{(A)_{k}}{}_{\ldots }\equiv T_{\ldots
A_{1}\ldots A_{k}\ldots }Q_{\ldots }{}^{A_{1}\ldots A_{k}}{}_{\ldots }, \nn
&& (\partial_A)^Q\equiv \partial_{A_1}\partial_{A_2}\ldots
\partial_{A_Q},\ \  (p_A)^Q\equiv p_{A_1}p_{A_2}\ldots p_{A_Q},
\notag
\end{eqnarray}
and so on.

We denote by $M_p(\ldots)$ the separately continuous
superantisymmetrical $p$-linear forms on
$(\mathbf D_{n})^p$. Thus, the arguments of these functionals are the
functions $f(z)$ of the form
\begin{equation}\label{dec}
f(z)=\sum_{k=0}^{n} f_{(\alpha)_k}(x)(\xi^\alpha)^k \in {\mathbf D}_n,
\quad f_{(\alpha)_k}(x)\in \D(\R^{n}).
\end{equation}
For any $f(z)\in {\mathbf D}_n$ we can define the support
$$ \mathrm{supp}(f)\stackrel {def} = \bigcup_{(\alpha)_k}
\mathrm{supp}(f_{(\alpha)_k}(x)).$$
For each set $V\subset \R^n$ we use the notation $z\bigcap V=\varnothing$
if $z=(x,\xi)$ and there exist some domain $U\subset \R^n$ such that
 $x\in U$ and $U\bigcap V=\varnothing$.

It can be easily proved
that such multilinear forms can be written in the integral form (see
\cite{SKT1}):
\begin{equation} M_p(f_1,\ldots,f_p)=\int dz_p\cdots
dz_1m_p(z_1,\ldots,z_p) f_1(z_1)\cdots f_p(z_p),\;p=1,2,... \label{3.1n}
\end{equation}
and
\begin{equation} M_p(z|f_1,\ldots,f_p)=\int dz_p\cdots
dz_1m_p(z|z_1,\ldots,z_p) f_1(z_1)\cdots
f_p(z_p),\;p=1,2,\,...\,\,.\label{3.2n}
\end{equation}
Let by definition
\[
\epsilon(M_p(f_1,\ldots,f_p))
=\epsilon_{m_p}+pn+
\epsilon(f_1)+\ldots+\epsilon(f_p).
\]
It follows from the properties of the forms $M_p$ that the corresponding
kernels $m_p$ have the following properties:
\begin{eqnarray}
&&\epsilon_{m_p}=pn+\epsilon_{M_p},\;\;\varepsilon_{m_p}=pn+
\varepsilon_{M_p},\;\;\epsilon_{m_p}=\varepsilon_{m_p}+p+1, \nonumber \\
&&m_p(*|z_1\ldots z_i,z_{i+1}\ldots z_p)=(-1)^{n} m_p(*|z_{1}\ldots
z_{i+1}^*,z_i^*\ldots z_p). \label{3.2an}
\end{eqnarray}
Here  $z^*=(x,\,-\xi)$ if $z=(x,\,\xi)$.

Introduce the space ${\cal M}_1\subset {C}^2(\mathbf D_n,\,\mathbf D_n')$
consisting of all 2-forms which can be locally represented as
\begin{eqnarray}
M_{2|2}^{1}(z|f,g) = \sum_{q = 0}^Q
m^{1(A)_q}(z|[(\partial^z_A)^qf(z)]g- (-1)^{\epsilon(f)\epsilon(g)}
[(\partial^z_A)^qg(z)]f),
\label{5.4d}
\end{eqnarray}
with locally constant $Q$ and
the space ${\cal M}_2\subset {C}^2(\mathbf D_n,\,\mathbf D_n')$
consisting of all 2-forms which can be locally represented as
\begin{eqnarray}
M_{2|2}^{2}(z|f,g) = \sum_{q =
0}^Q m^{2(A)_q}(z|[(\partial_A)^qf]g- (-1)^{\epsilon(f)\epsilon(g)}
(\partial_A)^qg]f)
\label{5.4h}
\end{eqnarray}
with locally constant $Q$,
where
$m^{1,2(A)_q}(z|\cdot)\in {C}^1(\mathbf D_n,\,\mathbf D_n')$.

The space ${\cal M}_0 = {\cal M}_1 \bigcap {\cal M}_2$ is called in
this paper the space of local bilinear forms. It consists of all the form,
which can be present as
$$
M_{2|\,\mathrm {loc}}(z|f,g) = \sum_{p,\,q =
0}^Q m^{(A)_q|(B)_p}(z)\left((\partial_A)^q f(z) \, (\partial_B)^p g (z) -
(-1)^{\epsilon(f)\epsilon(g)}
(\partial_A)^q  g(z)\, (\partial_B)^p f (z)
\right).
$$
Here $m^{(A)_q|(B)_p}\in D'\otimes \G^{n} $, and
the limit $Q$ is locally constant with respect to $z$.

The following low degree filtrations
$\oP_p$ and
$\oP_{p,q}$
of the polynomials we will use in what follows:
\begin{eqnarray*}
&& \mbox{\bf Definition. } \\
&&\oP_{p}=\{f(k)\in {\mathbf E_n} [k]:\, \exists g\in{\mathbf E_n}[\alpha,k]
\ \ \ f(\alpha k)=\alpha^p g(\alpha, k)\},        \\
&&\oP_{p,q}=\{f(k_1,k_l)\in {\mathbf E_n} [k_1,k_2]:\,
\exists g\in{\mathbf E_n}[\alpha,\beta,k_1,k_2]\ \
f(\alpha k_1,\beta k_2)=\alpha^p\beta^qg(\alpha,\beta,k_1,k_2)\}.
\end{eqnarray*}

Evidently, $\oP_{p,q}\subset \oP_{r,s}$ if $p \geq r$ and $q \geq s$. It is
clear also, that if $f\in \oP_{p,q}$ and $g\in \oP_{r,s}$ then
$fg\in\oP_{p+r,q+s}$. Analogous relations are valid for $\oP_p$.

\section{Cohomologies in the trivial representation}

In this section, we prove the theorem \ref{th1}.

\subsection{$H^1_{{\rm tr}}$}
\label{Htr1}

Let $
M_1(f)=\int dzm_1(z)f(z)
$.
Then the cohomology equation has the form
$0=
d_1^{{\rm tr}}M_1(f,g)=-M_1([f,g])=-\int dzm_1(z)[f(z),g(z)],
$
and hence
$
m_1(z)\frac{\overleftarrow{\partial}}{\partial z^A}\omega^{AB} \frac{\partial
}{\partial z^B}f(z)+2m_1(z)\Delta f(z)=0.$
Finally,
$m_1(z)=0$, i.e., $H^1_{{\rm tr}}=0$.

\subsection{$H^2_{{\rm tr}}$}

For the bilinear form
$
M_{2}(f,g)=\int dudzm_{2}(z,u)f(z)g(u)
$
the cohomology equation has the form
\begin{equation}
M_{2}([f,g],h)-(-1)^{\epsilon (g)\epsilon
(h)}M_{2}([f,h],g)-M_{2}(f,[g,h])=0.  \label{-2}
\end{equation}

Let
$
{\rm supp}(h)\bigcap \left[ {\rm supp}(f)\bigcup {\rm supp}(g)\right]
=\varnothing.
$
Then we have
$
\hat{M}_{2}([f,g],h)=0
$
(the hat means that the corresponding form (or kernel) is considered out of
the diagonal),
which imply $\hat{m}_{2}(z,u)=0$.
Thus we can represent $M_{2}(f,g)$ in the following form
(see \cite{SKT1})
\begin{eqnarray}
\!\!\!\!M_{2}(f,g)
\!\!\!&=&\!\!\!\sum_{k=0}^{K}\int dzm_{2}^{(A)_{k}}(z)\left(
(-1)^{\varepsilon (f)}(\partial _{A})^{k}f(z)\cdot g(z)-
(-1)^{\varepsilon
(g)+\epsilon (f)\epsilon (g)}(\partial
_{A})^{k}g(z)\cdot f(z)\right),\nn
\label{-1}
\end{eqnarray}
where upper limit $K$ is locally constant.

\proposition\label{proposition1}
{\it Summation in formula (\ref{-1}) is made over even }
$k$.

In particular, the highest degree $K$ of derivatives in
formula (\ref{-1}) is even, $K=2m$.

Indeed, let $K_{0}$ be the highest odd degree of derivatives. Then a
summands with $k=K_0$ in the second term is equal to
\begin{eqnarray*}
&&\int dz m_{2}^{(A)_{K_0}}(z)(-1)^{\varepsilon (f)\varepsilon
(g)}[(\partial _{A})^{K_0}g(z)]f(z)= \\
&=&-\int dz m_{2}^{(A)_{K_0}}(z)[(\partial
_{A})^{K_0}f(z)]g(z)+\sum_{k<K_{0}}\int dzw_{2}^{(A)_{k}}(z)g(z)(\partial
_{A})^{k}f(z)
\end{eqnarray*}
and terms with $k=K_0$ are canceled in Exp. (\ref{-1}).

The cohomology equation has the form
\begin{eqnarray}
&&\sum_{k=0}^{2m}\int dzm_{2}^{(A)_{k}}((-1)^{\varepsilon (f)+\varepsilon
(g)}\{((\partial _{A})^{k}[f,g])h+(-1)^{(\varepsilon (f)+\varepsilon
(g)+1)\varepsilon (h)}((\partial _{A})^{k}h)[f,g]\}+  \nonumber \\
&&+(-1)^{\varepsilon (g)\varepsilon (h)+\varepsilon (f)+\varepsilon
(g)}\{((\partial _{A})^{k}[f,h])g+(-1)^{\varepsilon (g)(\varepsilon
(f)+\varepsilon (h)+1)}((\partial _{A})^{k}g)[f,h]\}+  \nonumber \\
&&\,+(-1)^{\varepsilon (f)}\{((\partial _{A})^{k}f)[g,h]+(-1)^{\varepsilon
(f)[\varepsilon (g)+\varepsilon (h)+1]}((\partial _{A})^{k}[g,h])f\})=0.
\label{0}
\end{eqnarray}

Analogously to \cite{Zh},
take the functions in the form $f(z)\rightarrow e^{zp}f(z)$, $g(z)\rightarrow
e^{zq}g(z)$, $\ h(z) \rightarrow e^{-z(p+q)}h(z)$,
and consider the terms of the highest order in $p$ and $q$ which equals to
$2m+2$. Using the notation $\left\langle p,q\right\rangle \equiv
\sum_{A,B}(-1)^{\varepsilon_{A}}\omega^{AB}p_{A}q_{B}=\left\langle
q,p\right\rangle$ ($\varepsilon(\left\langle p,q\right\rangle) =1$) and,
introducing the generation function
\[
F_{m}(z,p)\equiv m_{2}^{(A)_{2m}}(z)(p_{A})^{2m},
\]
we obtain
\begin{equation}
\langle
p,p\rangle F_{m}(z,q)+\langle q,q\rangle F_{m}(z,p)+\langle p,q\rangle
\{F_{m}(z,p)+F_{m}(z,q)-F_{m}(z,p+q)\}=0
\label{1}
\end{equation}
Let $p=(v_{i},\theta _{i})$, $q=(y_{i},\zeta _{i})$.
Then we can rewrite (\ref{1}) in the form
\begin{equation}
2v\theta F_{m}(z,q)+2y\zeta F_{m}(z,p)+(x\zeta +\theta
y)\{F_{m}(z,p)+F_{m}(z,q)-F_{m}(z,p+q)\}=0  \label{1a}
\end{equation}

\proposition\label{propi}
{\it Let $n>1$, $m\geq 2$. Then
\be
F_{m}(z,p)=0.  \label{1b}
\ee
}
\begin{proof}
Let us differentiate Eq. (\ref{1a}) by $\partial _{q^{B}}\partial _{q^{A}}$
at $q=0$. We obtain
\begin{eqnarray}
&&\theta ^{i}\partial _{v^{j}}F_{m}(z,p)+\theta ^{j}\partial
_{v^{i}}F_{m}(z,p)=0\;\Longrightarrow \;\partial _{v^{i}}F_{m}(p)=2\theta
^{i}a_{1}(z,p), \nonumber\\
&&v^{i}\partial _{\theta ^{j}}F_{m}(z,p)-v^{j}\partial _{\theta
^{i}}F_{m}(z,p)=0\;\Longrightarrow \;\partial _{\theta
^{i}}F_{m}(z,p)=2v^{i}a_{2}(z,p), \nonumber\\
&&2\delta _{ij}F_{m}(z,p)+\theta ^{i}\partial _{\theta
^{j}}F_{m}(z,p)-v^{j}\partial _{v^{i}}F_{m}(z,p)=0\;\Longrightarrow
\nonumber
\end{eqnarray}
$$
\delta _{ij}F_{m}(z,p)=v^{j}\theta ^{i}a(z,p)=v^{i}\theta
^{j}a(z,p),\;a(z,p)=a_{1}(z,p)-a_{2}(z,p)
$$
Let $i=1$, $j=\bar{j}\neq 1$ ($j=(1,\bar{j})$). We have
\begin{equation*}
0=v^{\bar{j}}\theta ^{1}a(z,p)=v^{1}\theta ^{\bar{j}}a(z,p)\;\Longrightarrow
\theta ^{i}a(z,p)=0
\;\Longrightarrow F_m(z,p)=0.
\end{equation*}
\end{proof}

\proposition\label{propii}
{\it
Let $n>1$, $m=1$.
Then
\begin{equation}
F_{1}(z,p)=\frac{1}{2}b_{1}(z)\langle p,p\rangle,   \label{1c}
\end{equation}
i.e., $F_{1}(z,p)$ corresponds to the differential of some 1-form
(see Exp. (\ref{4})).
}

\begin{proof}
\[
F_{1}(z,p)=\frac{1}{2}p_{A}p_{B}P(z)_{AB},\;P_{AB}=\left(
\begin{array}{cc}
a_{ij} & b_{ij} \\
b_{ij}^{t} & c_{ij}
\end{array}
\right)
\]
From (\ref{1a}) it follows
\begin{eqnarray*}
&& 2v\theta a_{ij}+\theta ^{i}(a_{jk}v^{k}+b_{jk}\theta ^{k})+\theta
^{j}(a_{ik}v^{k}+b_{ik}\theta ^{k}) = 0\;\Longrightarrow b_{ik}=b_{1}\delta
_{ik},\;\;a_{ij}=0\;, \\
&& 2v\theta c_{ij}+v^{i}(v^{j}+c_{jk}\theta ^{k})-v^{j}(v^{i}+c_{ik}\theta
^{k})  = 0\;\Longrightarrow c_{ij}=0
\;\Longrightarrow
\;P(z)_{AB}=b_1
(-1)^{\varepsilon_{A}}\omega^{AB}.
\end{eqnarray*}
\end{proof}

\proposition\label{propiii}
{\it
Let $n>1$, $m=0$. Then $F_{0}(z,p)=0$.
}

\begin{proof}
Indeed, $F_0(z,p)=b_0(z)$,
and it follows from (\ref{1a}) that
\begin{equation}
(2v\theta +2y\zeta +v\zeta +\theta y)b_{0}(z)=0\;\Longrightarrow b_{0}(z)=0.
\label{1d}
\end{equation}
\end{proof}

\proposition\label{propiv}
{\it
Let $n=1$.
Then
\[
F(z,p)=F_{1|1}(z)v^2+F_{2|1}(z)v\theta+F_{2|2}(z)v^{3}\theta .
\]}

\begin{proof}{We have in the case under consideration
\[
F_m(z,p)=F_{1|m}(z)v^{2m}+F_{2|m}(z)v^{2m-1}\theta .
\]
Eq. (\ref{1a}) takes the form
\begin{eqnarray*}
&&2v\theta (F_{1|m}y^{2m}+F_{2|m}y^{2m-1}\zeta )+2y\zeta
(F_{1|m}v^{2m}+F_{2|m}v^{2m-1}\theta )+ \\
&&+v\zeta (F_{1|m}v^{2m}+F_{2|m}v^{2m-1}\theta
+F_{1|m}y^{2m}-F_{1|m}(v+y)^{2m}-F_{2|m}(v+y)^{2m-1}\theta )+ \\
&&\,+y\theta (F_{1|m}v^{2m}+F_{1|m}y^{2m}+F_{2|m}y^{2m-1}\zeta
-F_{1|m}(v+y)^{2m}-F_{2|m}(v+y)^{2m-1}\zeta )=0.
\end{eqnarray*}
So
\begin{eqnarray}
F_{1|m}[2vy^{2m}+v^{2m}y+y^{2m+1}-(v+y)^{2m}y] &=&0,  \label{2} \\
F_{2|m}[2vy^{2m-1}-2v^{2m-1}y-v^{2m}+v(v+y)^{2m-1}+y^{2m}-(v+y)^{2m-1}y]
&=&0.  \label{3}
\end{eqnarray}
Setting $y=v$ in Eq. (\ref{2}), we obtain
$
(4-2^{2m})v^{2m+1}F_{1|m}(z)=0,
$
such that $F_{1|m}(z)=0$ if $m \ne 1$.

For $m\geq 3$ Eq. (\ref {3}) takes the form
$
F_{2|m}(z)[2(m-2)v^{2m-1}y+O(v^{2m-2})]=0$ and so
$F_{2|m}(z)=0$ for $m\geq 3$.
Eq. (\ref{2}) for $m=1$ and Eq. (\ref{3}) for $m=1,2$ are identically
satisfied for arbitrary $F_{1|1}(z)$, $F_{2|1}(z)$, and $F_{2|2}(z)$.
}
\end{proof}

Note that
$F_{2|1}(z)v\theta$ corresponds to some coboundary, i.e.
to the differential of some 1-form.

Return to the complete form of cohomology equation.

It is useful to rewrite a trivial solution of
cohomology equation (\ref{-2}),
$$
M_{2|{\rm tr}}(f,g)=d_{1}^{{\rm tr}}M_{1}(f,g)=\int dzm_{1}(z)[f(z),g(z)],
$$
in the form (\ref{-1}),
\begin{eqnarray}
&&M_{2|{\rm tr}}(f,g)=-\int dz m_{1}(z)(-1)^{\varepsilon (f)}\left( [\Delta
f(z)]g(z)+(-1)^{\varepsilon (f)\varepsilon (g)}[\Delta g(z)]f(z)\right) -
\nonumber \\
&&\,-\int dzm_{1}(z)\overleftarrow{\Delta }(-1)^{\varepsilon
(f)}f(z)g(z).  \label{4}
\end{eqnarray}

Consider the case $n>1$. According to Eqs. (\ref{1b}), (\ref{1c})
and (\ref{4}), $M_{2}(f,g)$ can be represented in the form
$$
M_{2}(f,g)=d_{1}^{{\rm tr}}M_{1}(f,g)+M_{2}^{\prime
}(f,g),\;m_{1}(z)=-b_{1}(z),$$
$$M_{2}^{\prime }(f,g)=\int
dz m^{\prime 0}(z)(-1)^{\varepsilon (f)}f(z)g(z),\;
m^{\prime 0}(z)=m^0(z) +  m_1(z) \overleftarrow{\Delta}.
$$
The form $M_{2}^{\prime }(f,g)$ satisfies the cohomology equation (\ref
{-2}), and therefore, according to Eq. (\ref{1d}), $M_{2}^{\prime }(f,g)=0$.
Thus, we have proved the following proposition
 \proposition
{\it
Let $n>1$. Then the general solution of the cohomology equation (\ref
{-2}) has the form
\[
M_{2}(f,g)=d_{1}^{{\rm tr}}M_{1}(f,g).
\]
}

Now consider the case $n=1$. According to above consideration, in this case the
solution of the cohomology equation (\ref{-2}) can be represented in the
form
\begin{eqnarray*}
&&M_{2}(f,g)=M_{2|4}(f,g)+M_{2|2}(f,g)+M_{2|0}(f,g)+d_{1}^{{\rm tr}
}M_{1}(f,g),
\end{eqnarray*}
where
\begin{eqnarray*}
&&M_{2|4}(f,g)=\int dzm_{4}(z)(-1)^{\varepsilon (f)}\{[\partial
_{x}^{3}\partial _{\xi }f(z)]g(z)+(-1)^{\varepsilon (f)\varepsilon
(g)}[\partial _{x}^{3}\partial _{\xi }g(z)]f(z)\}, \\
&&M_{2|2}(f,g)=\int dz(-1)^{\varepsilon (f)}\{[m_{2|1}(z)\partial
_{x}^{2}f(z)]g(z)+(-1)^{\varepsilon (f)\varepsilon (g)}[m_{2|1}(z)\partial
_{x}^{2}g(z)]f(z)\}, \\
&&M_{2|0}(f,g)=\int dzm_{0}(z)(-1)^{\varepsilon (f)}f(z)g(z),\;\;
M_{1}(f)=\int dzm_{2|2}(z)f(z).
\end{eqnarray*}

Take the functions in the form $f(z)\rightarrow e^{zp}f(z)$, $
g(z)\rightarrow e^{zq}g(z)$, $\ h(z)\rightarrow e^{-z(p+q)}h(z)$, and consider
the terms of the sixth and fifth orders in $p$ and $q$ in cohomology
equation (\ref{-2}). Only $M_{2|4}(f,g)$ from $M_{2}(f,g)$ will take part to
such terms. The sixth order in $p$ and $q$ terms are cancel identically and
we obtain the following equation for the fifth order terms
\begin{eqnarray}
&&\partial _{x}m_{4}(z)(2v^{3}+3v^{2}y-3vy^{2}-2y^{3})\theta \zeta +
\nonumber \\
&&\,+m_{4}(z)\overleftarrow{\partial }_{\xi
}[(2v^{3}y+3v^{2}y^{2}+vy^{3})\theta +(v^{3}y+3v^{2}y^{2}+2vy^{3})\zeta ]=0.
\label{6}
\end{eqnarray}
It follows from Eq. (\ref{6}), that
$
\partial _{x}m_{4}(z)=m_{4}(z)\overleftarrow{\partial }_{\xi
}=0.$
So
$m_{4}(z)=\frac{1}{2}m_{1}={\rm const}.\;
$
Thus, we can write
\[
M_{2}(f,g)=m_{1}\mu _{1}(f,g)+\bar{M}_{2}(f,g)+d_{1}^{{\rm tr}}M_{1}(f,g),\;
\bar{M}_{2}(f,g)=M_{2|2}(f,g)+M_{2|0}(f,g),
\]
\begin{equation}
\mu _{1}(f,g)=\int dz(-1)^{\varepsilon (f)}\{\partial _{x}^{3}\partial _{\xi
}f(z)\}g(z),\;\epsilon _{\mu _{1}}=1.  \label{6a}
\end{equation}
The form $\mu _{1}(f,g)$ satisfies cohomology equation (\ref{-2}), such
that the form $\bar{M}_{2}(f,g)$ satisfies cohomology equation (\ref{-2})
too. Again, take the functions in the form $f(z)\rightarrow e^{zp}f(z)$, $
g(z)\rightarrow e^{zq}g(z)$, $\ h(z)\rightarrow e^{-z(p+q)}h(z)$, and consider
the terms of the fourth and third orders in $p$ and $q$ in cohomology
equation (\ref{-2}). Only $M_{2|2}(f,g)$ from $\bar{M}_{2}(f,g)$ will take
part to such terms. The fourth order in $p$ and $q$ terms are cancelled
identically and we obtain the equation for the third order terms
\begin{eqnarray}
&&[\partial_{x}^{2}m_{2|1}(z)y+\partial _{x}m_{2|1}(z)y(2v+y)]\theta
+[\partial _{x}^{2}m_{2|1}(z)v+\partial _{x}m_{2|1}(z)v(v+2y)]\zeta -
\nonumber \\
&&\,-2\partial _{x}\partial _{\xi }m_{2|1}(z)vy-\partial _{\xi
}m_{2|1}(z)vy(v+y)=0.  \label{7}
\end{eqnarray}
It follows from Eq. (\ref{7})
\[
\partial _{x}m_{2|1}(z)=m_{2|1}(z)\overleftarrow{\partial }_{\xi
}=0\;\Longrightarrow m_{2|1}(z)=\frac{1}{2}m_{2}={\rm const}.
\]
It is easy to prove that the form $M_{2|2}(f,g)$,
\begin{equation}
M_{2|2}(f,g)=m_{2}\mu _{2}(f,g),  \;\;
\mu _{2}(f,g)=\int dz(-1)^{\varepsilon (f)}
\{\partial_{x}^{2}f(z)\}g(z),\;\epsilon _{\mu _{2}}=0.
\label{8}
\end{equation}
satisfies cohomology equation (\ref{-2}). Therefore, the form
$M_{2|0}(f,g)$ satisfies cohomology equation (\ref{-2}) too. That means
$M_{2|0}(f,g)=0$.

Finally, we have proved the following proposition
\proposition
{\it
Let $n=1$. Then the general solution of cohomology equation (\ref{-2}) in
the trivial representation has the form
\[
M_{2}(f,g)=m_{1}\mu _{1}(f,g)+m_{2}\mu _{2}(f,g)+d_{1}^{{\rm tr}}M_{1}(f,g),
\]
where
\[
\mu_{1}(f,g)=\int dz(-1)^{\varepsilon(f)}
\{\partial_x^3\partial_{\xi}f(z)\}g(z), \;\;
\mu _{2}(f,g)=\int dz(-1)^{\varepsilon (f)}
\{\partial_{x}^{2}f(z)\}g(z).
\]
}

The forms $\mu _{1}(f,g)$ and $\mu _{2}(f,g)$ are
independent nontrivial cocycles. Indeed, suppose that the relation
\begin{equation}
m_{1}\mu _{1}(f,g)+m_{2}\mu _{2}(f,g)=-d_{1}^{{\rm tr}}M_{1}(f,g)=\int
dzm_{1}(z)[f(z),g(z)]  \label{10}
\end{equation}
is valid. Take the functions $f$\ and $g$ in the form $f(z)\rightarrow e^{zp}
f(z)$, $g(z)\rightarrow e^{-zp}g(z)$:
\begin{eqnarray*}
&&m_{1}\int dz(-1)^{\varepsilon (f)}\{[(v+\partial _{x})^{3}(\theta
+\partial _{\xi })f(z)]g(z)+
m_{2}\int dz(-1)^{\varepsilon (f)}\{[(v+\partial _{x})^{2}f(z)]g(z)= \\
&&=-\int dzm_{1}(z)\{f(z)(\overleftarrow{\partial}_A+
(-1)^{\varepsilon_A}p_A)\omega^{AB}(\partial_B-p_B)g(z)\}.
\end{eqnarray*}
Considering the terms of the fourth order in $p$, we obtain that
Eq. (\ref{10}) can be satisfied for $m_{1}=0$ only. Then, considering the terms of
the second order in $v$, we obtain that Eq. (\ref{10}) can be satisfied for $
m_{2}=0$ only.

\section{Zeroth and first adjoint cohomologies}

\subsection{$H^0_{\mathrm{ad}}$}

Let $
M_{0}(z)$ be 0-form. Then cohomology equation
$0=d_{0}^{\mathrm{ad}
}M_{0}(z|f_{1})=-[M_{0}(z),f_{1}(z)]
$
gives
$
M_{0}(z)=a=\mathrm{const.}
$
So $H^0_{\mathbf D_n'}\simeq H^0_{\mathbf E_n}\simeq \K$
and $H^0_{\mathrm{ad}} = 0$.

\subsection{$H^1_{\mathrm{ad}}$}\label{H^{1}}

Let
$M_{1}(z|f)=\int dum_{1}(z|u)f(u)
$ be 1-form. Then cohomology equation
$d_{1}^{\mathrm{ad}}M_{1}(z|f,g)=0$
gives
\begin{equation}
\lbrack M_{1}(z|f),g(z)]-(-1)^{\epsilon (f)
\epsilon
(g)}[M_{1}(z|g),f(z)]-M_{1}(z|[f,g])=0  \label{4.1_1}
\end{equation}

Let
$
z\bigcap \mathrm{supp}(f)=z\bigcap \mathrm{supp}(g)=\varnothing .
$
Then $\hat{M}_{1}(z|[f,,g])=0$ and
\begin{equation*}
\hat{m}_{1}(z|u)\overleftarrow{\partial }_{A}\omega _{\lambda }^{AB}\partial
_{B}f(z)+2\hat{m}_{1}(z|u)\Delta f(u)=0.
\end{equation*}
So $\hat{m}_{1}(z|u)=0$, and
\begin{equation*}
M_{1}(z|f)=\sum_{q=0}^{Q}t^{(B)_{q}}(z)(\partial
_{B})^{q}f(z),\;\varepsilon (t^{(B)_{q}}(z)(\partial _{B})^{q})=\varepsilon
_{M_{1}},
\end{equation*}

Let $f(z)=e^{zp}$, $g(z)=e^{zk}$ in some vicinity of $x$, and
\[
F(z,p)=\sum_{q=0}^{Q}t^{(B)_{q}}(z)(p_{B})^{q},\;\;
\epsilon(F(z,p))=\epsilon_{M_1}.
\]
The function $F(z,p)$ is polynomial in $p$ for every $z$
and a degree of this polynomial locally does not depend on $z$.

Then the cohomology equation acquires the form
\begin{equation}
(F(z,p)+F(z,k)-F(z,p+k))\left\langle p,k\right\rangle
+[F(z,k),zp]+[F(z,p),zk]=0.  \label{4.3}
\end{equation}

Consider the terms of highest order $Q+2$  in Eq. (\ref{4.3}).

Let $Q\geq 2$.
We obtain
\begin{equation}
(F_{Q}(z,p)+F_{Q}(z,k)-F_{Q}(z,p+k))\left\langle p,k\right\rangle =0,
\label{4.4}
\end{equation}
where
$
F_{Q}(z,p)=t^{(B)_{Q}}(z)(p_{B})^{Q}.
$
Acting on Eq. (\ref{4.4}) by the operator $\left. \frac{\overleftarrow{
\partial }}{\partial k^{A}}\frac{\overleftarrow{\partial }}{\partial k^{B}}
\right| _{k=0}$, we find
\begin{equation}
F_{Q}(z,p)\frac{\overleftarrow{\partial }}{\partial p_{A}}p_{C}\omega
^{CB}+(-1)^{\varepsilon _{A}\varepsilon _{B}}F_{Q}(z,p)\frac{\overleftarrow{
\partial }}{\partial p_{B}}p_{C}\omega ^{CA}=0.
\label{||}
\end{equation}
The general solution of (\ref{||}) has the form
\begin{equation}
F_{Q}(z,p)\frac{\overleftarrow{\partial }}{\partial p_{A}}
=t(z,p)(-1)^{\varepsilon _{C}}p_{C}\omega ^{CA},  \label{4.5}
\end{equation}
where $t(z,p)$ is some polynomial in $p$.
Using the property $$F_{Q}(z,p)
\frac{\overleftarrow{\partial }}{\partial p_{A}}\frac{\overleftarrow{
\partial }}{\partial p_{B}}-(-1)^{\varepsilon _{A}\varepsilon _{B}}F_{Q}(z,p)
\frac{\overleftarrow{\partial }}{\partial p_{B}}\frac{\overleftarrow{
\partial }}{\partial p_{A}}\equiv 0,$$
we obtain from Eq. (\ref{4.5})
\begin{equation*}
t(z,p)\frac{\overleftarrow{\partial }}{\partial p_{A}}p_{C}\omega
^{CB}-(-1)^{\varepsilon _{A}\varepsilon _{B}}t(z,p)\frac{\overleftarrow{
\partial }}{\partial p_{B}}p_{C}\omega ^{CA}=0
\end{equation*}
which implies
$t(z,p)\frac{\overleftarrow{\partial }}{\partial p_{A}}p_{A}=0$
and as a consequence $t(z,p)=t(z)$. So, we have
\begin{equation*}
Q=2,\;F_{2}(z,p)=\frac{1}{2}t(z)\left\langle p,p\right\rangle
,\;F(z,p)=t^{0}(z)+t^{A}(z)p_{A}+\frac{1}{2}t(z)\left\langle
p,p\right\rangle ,
\end{equation*}
and Eq. (\ref{4.3}) acquires the form
\begin{equation}
t^{0}(z)\left\langle p,k\right\rangle +[F(z,k),zp]+[F(z,p),zk]=0.
\label{4.6}
\end{equation}
Considering the terms of third order in $p$ and $k$ in Eq. (\ref{4.6}), we
obtain $t(z)\frac{\overleftarrow{\partial }}{\partial z^{A}}=0$ and so
$t(z)=t=\mathrm{const}$. Considering  the terms of first order in $p$ and
$k$ in Eq. (\ref{4.6}), we obtain
$t^{0}(z)\frac{\overleftarrow{\partial }}{\partial z^{A}}=0$ and so
$t^{0}(z)=t^{0}=\mathrm{const}$.

In such a way, Eq. (\ref{4.6}) is reduced to the equation
\begin{equation*}
t^{0}\omega ^{AB}+t^{B}(z)\frac{\overleftarrow{\partial }}{\partial z^{C}}
\omega ^{CA}(-1)^{\varepsilon _{A}+\varepsilon _{B}}+t^{A}(z)\frac{
\overleftarrow{\partial }}{\partial z^{C}}\omega ^{CB}(-1)^{\varepsilon
_{A}\varepsilon _{B}}=0
\end{equation*}
general solution of which has the form
\begin{equation*}
t^{A}(z)=-\frac{1}{2}t^{0}z^{A}+t_{1}(z)\frac{\overleftarrow{\partial }}{
\partial z^{B}}\omega ^{BA}.
\end{equation*}

Finally, we have obtained that general solution of the cohomology
equation (\ref{4.1_1}) has the form
\begin{equation*}
M_{1}(z|f)=t^{0}\eE_z f(z)+t\Delta f(z)+ d_0^{\mathrm {ad}}M_0(z|f),
\end{equation*}
where $M_0(z)=t_{1}(z)$.
Each summand in this expression satisfies the cohomology equation and
first two of them are nontrivial cocycles. Indeed, it is obvious that an
equation
\begin{equation*}
t^{0}\eE_z f(z)+t\Delta f(z)=[\phi (z),f(z)]
\end{equation*}
has solution for $t^{0}=t=0$ only.

Let us discuss the term $[t(z),f(z)]$. Is this expression a coboundary or
not? Analogously to \cite{SKT1},
the answer depends on the functional class $\mathcal{A}$ in which the
considered multilinear forms take their values.

1) $\mathcal{A}=\mathbf{D}_{n}^{\prime}$. In this case $t(z)\in
\mathbf{D}_{n}^{\prime}$ and the form $[t(z),f(z)]$ is exact.

2) $\mathcal{A}=\mathbf{E}_{n}$.
In this case
$t(z)\in \mathbf{E}_{n}$
and the form $[t(z),f(z)]$ is exact.

3) $\mathcal{A}=\mathbf{D}_{n}$. In this case the condition $
[t(z),f(z)]\in \mathbf{D}_{n}$ gives the restriction $t(z)\in
{\mathbf E_n }$ only, and the form $[t(z),f(z)]$ is exact if and only if $
t(z)\in \mathbf{D}_{n}\oplus {\cal C}_{\mathbf{E}_{n}}(\mathbf{D}_{n})$.
So the forms $[t(z),f(z)]$ are
independent nontrivial cocycles parametrized by the elements of factor-space
$\mathbf{E}_{n}/\mathbf{Z}_{n}$, where $\mathbf{Z}_{n}=\mathbf{D}_{n}
\oplus {\cal C}_{\mathbf{E}_{n}}(\mathbf{D}_{n})$.
Here ${\cal C}_{\mathbf{E}_{n}}(\mathbf{D}_{n})$ is a centralizer of
$\mathbf{D}_{n}$ in $\mathbf{E}_{n}$.
Evidently, ${\cal C}_{\mathbf{E}_{n}}(\mathbf{D}_{n})=\K$.

\section{Second adjoint cohomology}

For the bilinear form
\begin{equation*}
M_{2}(z|f,g)=\int dvdum_{2}(z|u,v)f(u)g(v)\in \mathcal{A}
\end{equation*}
the cohomology equation has the form
\begin{eqnarray}
&&d_{2}^{\mathrm{ad}}M_{2}(z|f,g,h)=-(-1)^{\epsilon (f)\epsilon (h)}\{(-1)^{\epsilon
(f)\epsilon (h)}[M_{2}(z|f,g),h(z)]+ \nonumber \\
&&+(-1)^{\epsilon (f)\epsilon (h)}M_{2}(z|[f,g],h)+
\mathrm{cycle}(f,g,h)\}=0.  \label{5.1}
\end{eqnarray}

\subsection{Nonlocal part}

\subsubsection{$n\geq 2$}\label{n geq 2}
Here we prove the following proposition:
\proposition\label{proptr>2}
{\it
Let $n\geq 2$. Then any adjoint 2-cocycle can be expressed in the form
\begin{equation*}
M_{2}(z|f,g)=M_{2|\mathrm{loc}}(z|f,g)+d_{1}^{\mathrm{ad}}M_{1|1}(z|f,g),
\end{equation*}
}
where $M_{2|\mathrm{loc}}\in {\cal M}_0$.

\begin{proof}{

Firstly, let us prove that
\begin{equation*}
M_{2}(z|f,g)=M_{2|2}(z|f,g)+d_{1}^{\mathrm{ad}}M_{1|1}(z|f,g),
\end{equation*}
}
where $M_{2|2}\in {\cal M}_2$. Let
\begin{equation*}
z\bigcap \left[ \mathrm{supp}(f)\bigcup \mathrm{supp}(g)\bigcup \mathrm{supp}
(h)\right] =\mathrm{supp}(f)\bigcap \left[ \mathrm{supp}(g)\bigcup \mathrm{
supp}(h)\right] =\varnothing .
\end{equation*}
We have $\hat{M}_{2}(z|f,[g,h])=0$, which implies
$\hat{m}_2(z|u,v)\overleftarrow{\partial^v}_A\omega^{AB}\partial_Bg(v)+
2\hat{m}_2(z|u,v)\Delta g(v)=0$
and so $\hat{m}_{2}(z|u,v)=0$.

As before, we can decompose
\begin{equation*}
M_{2}(z|f,g)=M_{2|1}(z|f,g)+M_{2|2}(z|f,g),
\end{equation*}
where $M_{2|1}\in {\cal M}_1$
and
$M_{2|2}\in {\cal M}_2$.

The proposition \ref{proposition1} can be applied to this case also,
and we can assume, that
the summation in the expression (\ref{5.4d}) for $M_{2|1}(z|f,g)$ is made over even $q$.

Let
\begin{equation*}
z\bigcap \left[ \mathrm{supp}(f)\bigcup \mathrm{supp}(g)\bigcup \mathrm{supp}
(h)\right] =\varnothing .
\end{equation*}
In this domain, $\hat {M}_{2|2}=0$ and we obtain
\begin{equation}
\hat{M}_{2|1}(z|[f,g],h)-(-1)^{\epsilon (h)\epsilon (g)}\hat{M}
_{2|1}(z|[f,h],g)-\hat{M}_{2|1}(z|f,[g,h])=0.  \label{5.2}
\end{equation}

This equation can be solved analogously to Eq. (\ref{-2}).

Using the proposition \ref{propi}
we can write (up to
local form which is included to $M_{2|2}(z|f,g)$)
\begin{eqnarray*}
M_{2|1}(z|f,g) \!\!\!&=&\!\!\!\frac{1}{2}\int dum^{1AB}(z|u)(-1)^{\varepsilon
(f)}\{[\partial _{A}^{u}\partial _{B}^{u}f(u)]g(u)+(-1)^{\varepsilon
(f)\varepsilon (g)}[\partial _{A}^{u}\partial _{B}^{u}g(u)]f(u)\}+ \\
&&\!\!\!\!\!\!+\int dum^{10}(z|u)(-1)^{\varepsilon (f)}f(u)g(u).
\end{eqnarray*}
Represent $M_{2|1}(z|f,g)$ in the form
\begin{eqnarray*}
&&\!\!\!\!\!\!M_{2|1}(z|f,g)=M_{2|1}^{\prime }(z|f,g)+d_{1}^{\mathrm{ad}
}M_{1|1}(z|f,g),\;M_{1|1}(z|f)=\int dum^{1}(z|u)f(u), \\
&&\!\!\!\!\!\!M_{2|1}^{\prime }(z|f,g)=\frac{1}{2}\int dum^{\prime
1AB}(z|u)(-1)^{\varepsilon (f)}\{[\partial _{A}^{u}\partial
_{B}^{u}f(u)]g(u)+(-1)^{\varepsilon (f)\varepsilon (g)}[\partial
_{A}^{u}\partial _{B}^{u}g(u)]f(u)\}+ \\
&&\!\!\!\!\!\!+\int dum^{\prime 10}(z|u)(-1)^{\varepsilon (f)}f(u)g(u), \\
&&\!\!\!\!\!\!m^{\prime 1AB}(z|u)=m^{1AB}(z|u)+(-1)^{\varepsilon _{A}}\omega
^{AB}m^{1}(z|u),\;(-1)^{\varepsilon _{A}}\omega _{AB}m^{\prime 1AB}(z|u)=0,
\\
&&\!\!\!\!\!\!m^{\prime10}(z|u)=m^{10}(z|u)+m^1(z|u)\overleftarrow{\Delta},
\;\; m^1(z|u)=\frac{1}{2n}(-1)^{\varepsilon_A}\omega_{AB}m^{1AB}(z|u),
\end{eqnarray*}
where $\omega_{AB}$ is defined by the relation $\omega_{AB}\omega^{BC}=\delta_A^C$.
The form $\hat{M}_{2|1}^{\prime }(z|f,g)$ satisfies Eq. (\ref{5.2}),
\begin{equation*}
\hat{M}_{2|1}^{\prime }(z|[f,g],h)-(-1)^{(\varepsilon (h)+1)(\varepsilon
(g)+1)}\hat{M}_{2|1}^{\prime }(z|[f,h],g)-\hat{M}_{2|1}^{\prime
}(z|f,[g,h])=0,
\end{equation*}
and, from the proposition \ref{propii}
it follows
$\hat{m}^{\prime 1AB}(z|u)p_{A}p_{B}=\frac{1}{2}\hat{b}_{1}(z|u)
\langle p,p\rangle$.
So $\hat{m}^{\prime 1AB}(z|u)=0$
and then $\hat{m}^{\prime 10}(z|u)=0$.

Thus, we have proved that
$M_{2}(z|f,g)=M_{2|2}(z|f,g)+d_{1}^{\mathrm{ad}}M_{1|1}(z|f,g)$.

Further, let
$
\left[ z\bigcup \mathrm{supp}(f)\bigcup \mathrm{supp}(g)\right] \bigcap
\mathrm{supp}(h)=\varnothing .
$
Then the cohomology equation (\ref{5.1}) gives
\begin{eqnarray*}
&&(-1)^{\epsilon (f)\epsilon
_{M_{2}}}[f(z),\sum_{q=0}^{Q}\int du\hat{m}^{2(A)_{q}}(z|u)(-1)^{
\varepsilon (g)}(\partial _{A}^{z})^{q}g(z)h(u)]- \\
&&-(-1)^{\epsilon (g)\epsilon _{M_{2}}+\epsilon
(f))}[g(z),\sum_{q=0}^{Q}\int du\hat{m}^{2(A)_{q}}(z|u)(-1)^{\varepsilon
(f)}(\partial _{A}^{z})^{q}f(z)h(u)]+ \\
&&\,+\sum_{q=0}^{Q}\int du\hat{m}^{2(A)_{q}}(z|u)(-1)^{\varepsilon
(f)+\varepsilon (g)}\{(\partial
_{A}^{z})^{q}[f(z),g(z)]\}h(u)=0,
\end{eqnarray*}
which implies
\begin{eqnarray}
&&(-1)^{\epsilon (f)\epsilon (g)+1}
\sum_{q=0}^{Q}[\hat{m}^{2(A)_{q}}(z|u)(\partial
_{A}^{z})^{q}g(z),f(z)]+  \notag \\
&&\,+\sum_{q=0}^{Q}[\hat{m}^{2(A)_{q}}(z|u)(\partial
_{A}^{z})^{q}f(z),g(z)]-\sum_{q=0}^{Q}\hat{m}^{2(A)_{q}}(z|u)(\partial
_{A}^{z})^{q}\{[f(z),g(z)]\}=0.  \label{5.3}
\end{eqnarray}

Take the functions in the form $f(z)= e^{zp}$, $g(z)=
e^{zq}$ in some vicinity of $x$,
and consider the terms of the highest order in $p$ and $q$ which
equals to $Q+2$. We have
\begin{eqnarray*}
\hat{m}^{2(A)_Q}(z|u)(p_A+k_A)^Q\left\langle p,k\right\rangle=0\;
\Rightarrow\hat{m}^{2(A)_Q}(z|u)=0\;\Rightarrow
\hat{m}^{2(A)_{q}}(z|u)=0,\;\forall q.
\end{eqnarray*}
Thus, proposition \ref{proptr>2} is proved.
\end{proof}

\subsubsection{$n=1$}

In this case each function $f(z)$ can be decomposed as
$f(z)=f_0(x)+\xi f_1(x)$.

\proposition\label{propadn=1}
{\it
Let $n=1$. Then any adjoint 2-cocycle can be expressed in the form
\begin{equation*}
M_{2}(z|f,g)=c_{1}m_{2|1}(z|f,g)+c_{2}m_{2|2}(z|f,g)+d_{1}^{\mathrm{ad}
}M_{1}(z|f,g)+M_{2\mathrm{loc}}(z|f,g),
\end{equation*}
where $c_i$ are constants.
}

The details of the proof can be found in Appendix \ref{appproof}.

\subsection{Local part}\label{subsection5.2}

Consider the local 2-form
\begin{eqnarray*}
&&M_{2|\mathrm{loc}}(z|f,g)=\sum_{a,b=0}^N
(-1)^{\varepsilon(f)(|\varepsilon_B|_{1,b}+1)}m^{(A)_a|(B)_b}(z)
[(\partial_A^z)^af(z)](\partial_B^z)^bg(z), \\
&&m^{(B)_q|(A)_p}=(-1)^{|\varepsilon_A|_{1,p}|\varepsilon_B|_{1,q}}
m^{(A)_p|(B)_q},\;\varepsilon(m^{(A)_p|(B)_q}(\partial_A^z)^a
(\partial_B^z)^b)=\varepsilon(M_{2}).
\end{eqnarray*}

The cohomology equation (\ref{5.1}) reduces to the equation
\begin{eqnarray}
&&\!\!\!\!\!\!d_{2}^{\mathrm{ad}}M_{2|\mathrm{loc}}(z|f,g)=0=  \notag \\
&&\!\!\!\!\!\!\,=
-(-1)^{\epsilon (f)(\epsilon (g)+\epsilon
(h))}[\sum_{a,b=0}^{N}(-1)^{\varepsilon (g)(|\varepsilon
_{B}|_{1,b}+1)}m^{(A)_{a}|(B)_{b}}(z)\{(\partial
_{A}^{z})^{a}g(z)\}(\partial _{B}^{z})^{b}h(z),f(z)]+  \notag \\
&&\!\!\!\!\!\!+(-1)^{\epsilon (g)\epsilon
(h)}[\sum_{a,b=0}^{N}(-1)^{\varepsilon (f)(|\varepsilon
_{B}|_{1,b}+1)}m^{(A)_{a}|(B)_{b}}(z)\{(\partial
_{A}^{z})^{a}f(z)\}(\partial _{B}^{z})^{b}h(z),g(z)]-  \notag \\
&&\!\!\!\!\!\!-[\sum_{a,b=0}^{N}(-1)^{\varepsilon (f)(|\varepsilon
_{B}|_{1,b}+1)}m^{(A)_{a}|(B)_{b}}(z)\{(\partial
_{A}^{z})^{a}f(z)\}(\partial _{B}^{z})^{b}g(z),h(z)]-  \notag \\
&&-\sum_{a,b=0}^{N}(-1)^{(\varepsilon (f)+\varepsilon (g)+1)(|\varepsilon
_{B}|_{1,b}+1)}m^{(A)_{a}|(B)_{b}}(z)\{(\partial
_{A}^{z})^{a}[f(z),g(z)]\}(\partial _{B}^{z})^{b}h(z)+  \notag \\
&&\!\!\!\!\!\!+(-1)^{\epsilon (g)(\epsilon
(h)}\sum_{a,b=0}^{N}(-1)^{(\varepsilon (f)+\varepsilon
(h)+1)(|\varepsilon _{B}|_{1,b}+1)}m^{(A)_{a}|(B)_{b}}(z)[(\partial
_{A}^{z})^{a}\{[f(z),h(z)]\}(\partial _{B}^{z})^{b}g(z)+  \notag \\
&&\!\!\!\!\!\!+\sum_{a,b=0}^{N}(-1)^{\varepsilon (f)(|\varepsilon
_{B}|_{1,b}+1)}m^{(A)_{a}|(B)_{b}}(z)\{(\partial
_{A}^{z})^{a}f(z)\}\{(\partial _{B}^{z})^{b}[g(z),h(z)]\}.  \label{6.1}
\end{eqnarray}

It is useful to give the expression for the local form, which is a
coboundary of a local 1-form
$T_{1|\mathrm{loc}}(z|f)=\sum_{a=0}^{K}t^{(A)_{p}}(z)(\partial
_{A}^{z})^{a}f(z)$, $\varepsilon(t^{(A)_a}(\partial_A^z)^a)=0$:
\begin{eqnarray}
&&M_{2|\mathrm{triv}}(z|f,g)=d_{1}^{\mathrm{ad}}T_{1|\mathrm{loc}
}(z|f,g)=[\sum_{a=0}^{K}t^{(A)_{p}}(z)(\partial _{A}^{z})^{a}f(z),g(z)]- \nn
&&-(-1)^{\epsilon (f)\epsilon(g)}[\sum_{a=0}^Kt^{(A)_p}(z)
(\partial_A^z)^ag(z),f(z)]-\sum_{a=0}^Kt^{(A)_p}(z)(\partial_A^z)^a
[f(z),g(z)].\label{locloc}
\end{eqnarray}
Let the functions have the form $f(z)= e^{zp}$, $g(z)=e^{zq}$,
$h(z)=e^{zr}$ in some vicinity of $x$.
Then Eq. (\ref{6.1}) acquires the form
\begin{eqnarray}
&&\Phi (z,p,q,r)\left\langle p,q\right\rangle +\Phi (z,q,r,p)\left\langle
q,r\right\rangle +\Phi (z,r,p,q)\left\langle r,p\right\rangle -  \notag \\
&&\,-[F(z,p,q),zr]-[F(z,q,r),zp]-[F(z,r,p),zq]=0,  \label{6.2}
\end{eqnarray}
where
\begin{eqnarray}
&&\Phi (z,p,q,r)=\Phi (z,q,p,r)=F(z,p+q,r)-F(z,p,r)-F(z,q,r),  \\
&&F(z,p,q)=F(z,q,p)=
\sum_{a,b=0}^{N}m^{(A)_{a}|(B)_{b}}(z)(p_{A})^{a}(q_{B})^{b}.  \notag
\end{eqnarray}
The function $F(z,p,q)$ is polynomial in $p$, $q$ for every $z$
and a degree of this polynomial locally does not depend on $z$.

\proposition
{\it Up to coboundary
\begin{eqnarray*}
M_{2|\mathrm{loc}}(z|f,g)=c_{3}m_{2|3}(z|f,g)+c_{4}m_{2|4}(z|f,g)+
M_{2|\mathrm{loc}}^{\prime }(z|f,g),
\end{eqnarray*}
where
$$M_{2|\mathrm{loc}}^{\prime }(z|e^{zp},e^{zq})e^{-z(p+q)}\in \oP_{1,1}.$$
}
\begin{proof}

Let $r=0$. Then Eq. (\ref{6.2}) reduced to
$$
\{F_{0}(z,p)+F_{0}(z,q)-F_{0}(z,p+q)\}\left\langle p,q\right\rangle
+[F_{0}(z,p),zq]+[F_{0}(z,q),zp]=0,
$$
where
$$
F_{0}(z,p)=F(z,p,0)=\sum_{a=0}^{N}m^{(A)_{a}|0}(z)(p_{A})^{a}.
$$
This equation coincides with the equation (\ref{4.3}) and its solution has
the following form
\begin{eqnarray*}
m^{0|0}(z) &=&t^{0},\;m^{A|0}(z)=-\frac{1}{2}t^{0}z^{A}+t_{1}(z)\frac{
\overleftarrow{\partial }}{\partial z^{B}}\omega ^{BA},\;m^{AB|0}(z)\partial
_{A}\partial _{B}=t\Delta , \\
m^{(A)_{a}|0}(z) &=&0,\;a\geq 3,\;t^{0}=\mathrm{const},\;t=\mathrm{const}
,\;\epsilon (t_{1}(z))=\epsilon _{M_{2}}+1.
\end{eqnarray*}
Let us note that

\begin{enumerate}

\item
\begin{eqnarray}
&&\,(-1)^{\varepsilon (f)}f(z)g(z)-\frac{1}{2}(-1)^{\varepsilon
(f)}\{N_{z}f(z)\}g(z)-\frac{1}{2}(-1)^{\varepsilon (f)}f(z)N_{z}g(z)=  \notag
\\
&&\,=m_{2|3}(z|f,g)+d_{1}^{\mathrm{ad}}T_{11}(z|f,g)-(-1)^{\varepsilon
(f)}\{N_{\xi }f(z)\}N_{\xi }g(z)-\frac{\left\langle z,z\right\rangle }{4}[
f(z),g(z)],  \notag
\end{eqnarray}
where $N_z$ is Euler operator and
$T_{11}(z|f)=\frac{\left\langle z,z\right\rangle }{4}f(z)$.

\item
\begin{eqnarray*}
&&(-1)^{\varepsilon(f)}\{\Delta f(z)\}g(z)+f(z)\Delta g(z)= \\
&&=m_{2|4}(z|f,g)+\frac{1}{2}(-1)^{\varepsilon(f)}\{\Delta f(z)\}N_{z}g(z)+
\frac{1}{2}\{N_{z}f(z)\}\Delta g(z),
\end{eqnarray*}

\item
\begin{eqnarray*}
&&(-1)^{\varepsilon(f)}m^{A|(B)_0}(z)\frac{\partial f(z)}{\partial z^A}g(z)+
(-1)^{\varepsilon(f)(\varepsilon_A+1)}m^{(B)_0|A}(z)f(z)\frac{
\partial g(z)}{\partial z^{A}}= \\
&&=(-1)^{\varepsilon (f)}[t_{1}(z),f(z)]g(z)+(-1)^{\varepsilon
(f)+\varepsilon (f)\varepsilon (g)}[t_{1}(z),g(z)]f(z)= \\
&&=d_{1}^{\mathrm{ad}}T_{12}(z|f,g)-t_{1}(z)[f(z),g(z)],
\;\; T_{12}(z|f)=t_{1}(z)f(z),\;\varepsilon _{T_{1}}=\varepsilon (t_{1}).
\end{eqnarray*}
\end{enumerate}
Choosing $t^{(A)_{0}}(z)=t_{1}(z)$ in Exp. (\ref{locloc}), we obtain (up to coboundary)
\begin{eqnarray*}
&&M_{2|\mathrm{loc}}(z|f,g)=c_{3}m_{2|3}(z|f,g)+c_{4}m_{2|4}(z|f,g)+M_{2|
\mathrm{loc}}^{\prime }(z|f,g), \\
&&M_{2|\mathrm{loc}}^{\prime }(z|f,g)=\sum_{a,b=1}^{N}(-1)^{\varepsilon
(f)(|\varepsilon _{B}|_{1,b}+1)}m^{(A)_{a}|(B)_{b}}(z)[(\partial
_{A})^{a}f(z)](\partial _{B})^{b}g(z),\;c_{3}=t^{0},\;c_{4}=t.
\end{eqnarray*}
\end{proof}

\proposition
{\it Up to coboundary
\begin{eqnarray*}
M_{2|\mathrm{loc}}(z|f,g)=c_{3}m_{2|3}(z|f,g)+c_{4}m_{2|4}(z|f,g)+
M_{2|\mathrm{loc}}^{\prime\prime }(z|f,g),
\end{eqnarray*}
where
$$M_{2|\mathrm{loc}}^{\prime\prime }(z|e^{zp},e^{zq})e^{-z(p+q)}\in \oP_{2,2}.$$
}
\begin{proof}
Considering in Eq. (\ref{6.2}) the linear in $r$ terms, we obtain
\begin{eqnarray}
&&[F^{A}(z,p+q)-F^{A}(z,p)-F^{A}(z,q)]\left\langle p,q\right\rangle - \\
&&-(-1)^{\varepsilon _{A}}F(z,p,q)\overleftarrow{\partial ^{z}}_{C}\omega
^{CA}-[F^{A}(z,q),zp]-[F^{A}(z,p),zq]=0,
\label{kru}
\\
&&F^{A}(z,p)=\left. F(z,p,q)\overleftarrow{\partial }_{q^{A}}\right|
_{q=0}=\sum_{a=1}^{N}m^{(B)_{a}|A}(z)(p_{B})^{a}.\notag
\end{eqnarray}

The linear in $q$ terms in (\ref{kru}) give:
\begin{eqnarray}
(-1)^{\varepsilon _{A}+\varepsilon _{B}+\varepsilon _{A}\varepsilon
_{B}}F^{B}(z,p)\overleftarrow{\partial ^{z}}_{C}\omega
^{CA}
+(-1)^{\varepsilon _{B}+\varepsilon _{A}\varepsilon _{B}}m^{A|B}
\overleftarrow{\partial ^{z}}_{C}\omega ^{CD}p_{D}+F^{A}(z,p)\overleftarrow{
\partial ^{z}}_{C}\omega ^{CB}=0.
\nn
\label{kru1}
\end{eqnarray}

The linear in $p$ terms in (\ref{kru1}) give:
\begin{eqnarray}
&&(-1)^{\varepsilon _{A}+\varepsilon _{B}+\varepsilon _{B}\varepsilon
_{C}}m^{A|B}(z)\overleftarrow{\partial }_{D}\omega ^{DC}+(-1)^{\varepsilon
_{A}+\varepsilon _{C}+\varepsilon _{A}\varepsilon _{B}}m^{C|A}(z)
\overleftarrow{\partial }_{C}\omega ^{CB}+  \notag \\
&&\,+(-1)^{\varepsilon _{B}+\varepsilon _{C}+\varepsilon _{A}\varepsilon
_{C}}m^{B|C}(z)\overleftarrow{\partial }_{C}\omega ^{CA}=0.  \label{6.3}
\end{eqnarray}
Multiply Eq. (\ref{6.3}) by $(-1)^{\varepsilon _{A}\varepsilon
_{C}}\varkappa _{A}\varkappa _{B}\varkappa _{C}$, $\varepsilon (\varkappa
_{A})=\varepsilon _{A}$, from the right, we obtain
\[
m(z,\varkappa)\overleftarrow{d}=0,\;
m(z,\varkappa )=m^{A|B}(z)\varkappa_{A}\varkappa _{B}, \;\;
\overleftarrow{d}=\overleftarrow{\partial}_C\omega^{CA}\varkappa_A,\;
\overleftarrow{d}\overleftarrow{d}=0,
\]
from what it follows in the standard way
\begin{equation*}
m(z,\varkappa )=t(z,\varkappa )\overleftarrow{d},\;t(z,\varkappa
)=2t^{A}(z)\varkappa _{A}\;\Longrightarrow
\end{equation*}
\begin{equation*}
m^{A|B}(z)=(-1)^{\varepsilon _{A}+\varepsilon _{A}\varepsilon _{B}}t^{A}(z)
\overleftarrow{\partial }_{C}\omega ^{CB}+(-1)^{\varepsilon _{B}}t^{B}(z)
\overleftarrow{\partial }_{C}\omega ^{CA}.
\end{equation*}

The nonlinear in $p$ terms in (\ref{kru1}) give:
\begin{eqnarray}
&&\,(-1)^{\varepsilon _{A}}F^{\prime A}(z,p)\overleftarrow{\partial ^{z}}
_{C}\omega ^{CB}+(-1)^{\varepsilon _{B}+\varepsilon _{A}\varepsilon
_{B}}F^{\prime B}(z,p)\overleftarrow{\partial ^{z}}_{C}\omega ^{CA}=0,
 \label{6.4}\\
&&F^{\prime A}(z,p)=\sum_{a=2}^{N}m^{(B)_{a}|A}(z)(p_{B})^{a}.\notag
\end{eqnarray}
Multiplying Eq. (\ref{6.4})  from the right
 by $\varkappa _{B}\varkappa _{A}$
we obtain
$
F^{\prime }(z,p,\varkappa )\overleftarrow{d}=0,\;F^{\prime }(z,p,\varkappa
)=F^{\prime A}(z,p)\varkappa _{A},
$
from what it follows in the standard way
\begin{equation*}
F^{\prime }(z,p,\varkappa )=t^{\prime }(z,p)\overleftarrow{d},\;t^{\prime
}(z,p)=\sum_{a=2}^{N}t^{(A)_{p}}(z)(p_{A})^{a}\;\Longrightarrow
\end{equation*}
\begin{equation*}
F^{\prime A}(z,p)=t^{\prime }(z,p)\overleftarrow{\partial ^{z}}_{C}\omega
^{CA}.
\end{equation*}

So, the terms in $M_{2|\mathrm{loc}}^{\prime }(z|f,g)$ proportional to $
\partial f$, or $\partial g$, or $\partial f\partial g$ have the following
structure
\begin{eqnarray*}
&&\sum_{a=2}^{N}t^{(A)_{a}}\overleftarrow{\partial ^{z}}_{C}\omega
^{CB}(-1)^{(\varepsilon _{B}+1)|\varepsilon _{A}|_{2,N}+\varepsilon
(f)(\varepsilon _{B}+1)}[(\partial _{A})^{a}f(z)]\partial _{B}g(z)+ \\
&&\sum_{a=2}^{N}t^{(B)_{a}}\overleftarrow{\partial ^{z}}_{C}\omega
^{CA}(-1)^{|\varepsilon _{B}|_{2,N}+\varepsilon (f)(\varepsilon
_{B}|_{2,N}+1)}\partial _{A}f(z)[(\partial _{B})^{a}g(z)]+ \\
&&+(-1)^{\varepsilon (f)(\varepsilon _{B}+1)}\{(-1)^{\varepsilon
_{A}+\varepsilon _{A}\varepsilon _{B}}t^{A}(z)\overleftarrow{\partial }
_{C}\omega ^{CB}+(-1)^{\varepsilon _{B}}t^{B}(z)\overleftarrow{\partial }
_{C}\omega ^{CA}\}[\partial _{A}f(z)]\partial _{B}g(z) \\
&=&d_{1}^{\mathrm{ad}}T_{1}(z|f,g)+\mathrm{more},\;T_{1}(z|f)=
\sum_{a=1}^{N}t^{(A)_{a}}(\partial _{A})^{a}f(z),
\end{eqnarray*}
and ``more'' means the terms proportional to $\partial ^{a}f\partial ^{b}g\;$
with $a,b\geq 2$. Thus, we obtain (up to coboundary)
\begin{eqnarray*}
&&M_{2|\mathrm{loc}}(z|f,g)=c_{3}m_{2|3}(z|f,g)+c_{4}m_{2|4}(z|f,g)+M_{2|
\mathrm{loc}}^{\prime \prime }(z|f,g), \\
&&\,M_{2|\mathrm{loc}}^{\prime \prime
}(z|f,g)=\sum_{a,b=2}^{N}(-1)^{\varepsilon (f)(|\varepsilon
_{B}|_{1,b}+1)}m^{(A)_{a}|(B)_{b}}(z)[(\partial _{A})^{a}f(z)](\partial
_{B})^{b}g(z),
\end{eqnarray*}
\end{proof}

The cohomology equation takes the form of Eq. (\ref{6.2}) where now
\begin{equation*}
F(z,p,q)=F(z,q,p)\in \oP_{2,\,2}\,.
\end{equation*}

\proposition
{\it
If $F(z,p,q)$ is solution of Eq. (\ref{6.2}) and
$F(z,p,q)\in \oP_{2,\,2}$ then $F(z,p,q)$ does not depend on $z$.
}

\begin{proof}
Considering in Eq. (\ref{6.2}) the linear in $r$ terms, we obtain
\begin{equation*}
F(z,p,q)\overleftarrow{\partial ^{z}}_{C}=0\;\Longrightarrow
\;m^{(A)_{a}|(B)_{b}}(z)=m^{(A)_{a}|(B)_{b}}=\mathrm{const}.
\end{equation*}
\end{proof}

The cohomology equation takes the form
\begin{eqnarray}
&&\Phi (p,q,r)\left\langle p,q\right\rangle +\Phi (q,r,p)\left\langle
q,r\right\rangle +\Phi (r,p,q)\left\langle r,p\right\rangle =0,  \label{6.5}
\\
&&\Phi (p,q,r)=\Phi (q,p,r)=F(p+q,r)-F(p,r)-F(q,r),  \notag \\
&&F(p,q)=F(q,p)\in\oP_{2,\,2}.
\notag
\end{eqnarray}

Acting on Eq. (\ref{6.5}) by the operator
$\left.\overleftarrow{\partial }_{r_A}\overleftarrow{
\partial }_{r_B}\right| _{r=0} $ we obtain

\begin{eqnarray}
&&F(p,q)\left( \overleftarrow{L}_{p}^{AB}+\overleftarrow{L}_{q}^{AB}\right)
=[F^{AB}(p+q)-F^{AB}(p)-F^{AB}(q)]\left\langle p,q\right\rangle ,
\label{6.6}
\end{eqnarray}
where
\begin{eqnarray}
F^{AB}(p)&=&
\left.
F(p,q)\overleftarrow{\partial }_{q_A}\overleftarrow{
\partial }_{q_B}
\right| _{q=0},  \notag \\
\overleftarrow{L}_{p}^{AB}
&=&
\overleftarrow{\partial }_{p_A}p_{C}
\omega^{CB}+(-1)^{\varepsilon _{A}\varepsilon _{B}}
\overleftarrow{\partial }
_{p_B}p_{C}\omega ^{CA},\;\varepsilon \left( \overleftarrow{L}_{p}^{AB}\right)
=\varepsilon _{A}+\varepsilon _{B}+1.
\end{eqnarray}

It follows from Eq. (\ref{6.6}) that
\begin{eqnarray}
&&[(-1)^{\varepsilon _{A}+\varepsilon _{B}}F^{AB}(p)]\overleftarrow{L}
_{p}^{CD}-(-1)^{(\varepsilon _{C}+\varepsilon _{D}+1)(\varepsilon
_{A}+\varepsilon _{B}+1)}[(-1)^{\varepsilon _{C}+\varepsilon
_{D}}F^{CD}(p)]\overleftarrow{L}_{p}^{AB}  \notag \\
&&\,=-\{[(-1)^{\varepsilon _{A}+\varepsilon _{D}}F^{AD}(p)]\omega
^{BC}+(-1)^{\varepsilon _{C}\varepsilon _{D}}[(-1)^{\varepsilon
_{A}+\varepsilon _{C}}F^{AC}(p)]\omega ^{BD}+  \notag \\
&&+(-1)^{\varepsilon _{A}\varepsilon _{B}}[(-1)^{\varepsilon
_{B}+\varepsilon _{D}}F^{BD}(p)]\omega ^{AC}+(-1)^{\varepsilon
_{A}\varepsilon _{B}+\varepsilon _{C}\varepsilon _{D}}[(-1)^{\varepsilon
_{B}+\varepsilon _{C}}F^{BC}(p)]\omega ^{AD}\},  \label{6.7}
\end{eqnarray}


\font\frtnfr=eufm10   scaled\magstep1
\font\twlfr=eufm10
\font\tenfr=eufm10
\font\egtfr=eufm8
\font\sixfr=eufm6

\newfam\frfam
\textfont\frfam=\frtnfr
\scriptfont\frfam=\twlfr
\scriptscriptfont\frfam=\tenfr

\def\fr{\fam\frfam}
\font\frtnopen=msbm10  scaled\magstep2
\font\twlopen=msbm10
\font\tenopen=msbm10
\font\egtopen=msbm8
\font\sixopen=msbm6

\newfam\openfam
\textfont\openfam=\frtnopen
\scriptfont\openfam=\twlopen
\scriptscriptfont\openfam=\tenopen

\def\open{\fam\openfam}
\font\frtnsf = cmss12 scaled\magstep1
\font\twlsf = cmss10
\font\tensf = cmss9
\font\egtsf = cmss8
\font\sixsf = cmss8
\newfam\Scfam
\textfont\Scfam = \frtnsf
\scriptfont\Scfam = \twlsf
\scriptscriptfont\Scfam = \tensf
\def\Sc{\fam\Scfam}


The operators $\overleftarrow{L}_{p}^{AB}$ constitute
Lie superalgebra ${\fr pe}$.

\proposition\label{eqSPE}
{\it
The solution of Eq. (\ref{6.7})
satisfying the condition $F^{AB}(p) \in \oP_{2}$
has the form
$$
F^{AB}(p) =-(-1)^{\varepsilon _{A}+\varepsilon _{B}}\varphi (p)
\overleftarrow{L}_{p}^{AB}+c(-1)^{\varepsilon _{A}}
\left\langle p,p\right\rangle \omega^{AB},
$$
where $\varphi (p)$ is arbitrary polynomial with property $\varphi
(p)\in \oP_{2}$ and $c$ is constant.
}

The proof of this proposition can be found in Appendix \ref{SPE}.

Then, it follows from Eq. (\ref{6.6})
\begin{equation*}
\left\{ F(p,q)-[\varphi (p+q)-\varphi (p)-\varphi (q)]\left\langle
p,q\right\rangle \right\} \left( \overleftarrow{L}_{p}^{AB}+\overleftarrow{L}
_{q}^{AB}\right) =0,
\end{equation*}
and one can easily obtain
\begin{eqnarray*}
&&F(p,q)=[\varphi (p+q)-\varphi (p)-\varphi (q)+c_{5}\left\langle
p,p\right\rangle +c_{5}\left\langle q,q\right\rangle + \\
&&\,+c_{6}\left\langle p,p\right\rangle \left\langle q,q\right\rangle
]\left\langle p,q\right\rangle +c_{7}\left\langle p,p\right\rangle
\left\langle q,q\right\rangle
\end{eqnarray*}
($c_i$ are constants), which gives (taking in account the condition $F(p,q)=F(q,p)$)
\begin{equation*}
F(p,q)=[\varphi (p+q)-\varphi (p)-\varphi (q)+c_{5}\left\langle
p,p\right\rangle +c_{5}\left\langle q,q\right\rangle ]\left\langle
p,q\right\rangle.
\end{equation*}

Then Eq. (\ref{6.5}) takes the form (all terms including the function $\varphi $
being cancelled identically)
\begin{eqnarray*}
&&c_{5}[\left\langle p,p\right\rangle \left\langle q,r\right\rangle
\left\langle p,q\right\rangle +\left\langle q,q\right\rangle \left\langle
r,p\right\rangle \left\langle p,q\right\rangle +\left\langle
q,q\right\rangle \left\langle r,p\right\rangle \left\langle q,r\right\rangle
+ \\
&&\,+\left\langle r,r\right\rangle \left\langle p,q\right\rangle
\left\langle q,r\right\rangle +\left\langle r,r\right\rangle \left\langle
p,q\right\rangle \left\langle r,p\right\rangle +\left\langle
p,p\right\rangle \left\langle q,r\right\rangle \left\langle r,p\right\rangle
]=0,
\end{eqnarray*}
which implies
$c_{5}=0\ \forall n\geq 1$,
such that we have
\begin{equation*}
F(p,q)=[\varphi (p+q)-\varphi (p)-\varphi (q)]\left\langle p,q\right\rangle .
\end{equation*}

Thus, the form $M_{2|\mathrm{loc}}^{\prime \prime }(z|f,g)$ is equal to
\begin{eqnarray*}
&&M_{2|\mathrm{loc}}^{\prime \prime
}(z|f,g)=[T_{1}(z|f),g]-(-1)^{(\varepsilon (f)+1)(\varepsilon
(g)+1)}[T_{1}(z|g),f]- \\
&&\,-[T_{1}(z|[f,g])=d_{1}^{\mathrm{ad}}T_{1}(z|f,g),\;T_{1}(z|f)=\varphi
(\partial_z )f(z).
\end{eqnarray*}

Finally, we obtained
\begin{eqnarray}
&&M_{2}(z|f,g)=\delta _{1n}[c_{1}m_{2|1}(z|f,g)+c_{2}m_{2|2}(z|f,g)]+  \notag
\\
&&\,+c_{3}m_{2|3}(z|f,g)+c_{4}m_{2|4}(z|f,g)+d_{1}^{\mathrm{ad}}M_{1}(z|f,g),
\label{6.7a}
\end{eqnarray}
the expressions for the forms $m_{2|1}(z|f,g)$, $m_{2|2}(z|f,g)$, $
m_{2|3}(z|f,g)$ and $c_{4}m_{2|4}(z|f,g)$ are given by Eqs. (\ref{5.2.7}), (
\ref{5.2.8}), (\ref{6.3a}) and (\ref{6.3b}), respectively.

\subsection{Independence and non triviality}

All forms $m_{2|a}(z|f,g)$, $a=1,2,3,4$, are independent nontrivial
cocycles for $n=1$, and forms $m_{2|a}(z|f,g)$, $a=3,4$, are independent
nontrivial cocycles for $n\geq 2$.

\subsubsection{$n=1$}

Let
\begin{eqnarray}
&&c_{1}m_{2|1}(z|f,g)+c_{2}m_{2|2}(z|f,g)+c_{3}m_{2|3}(z|f,g)+c_{4}m_{2|5}(z|f,g)=
d_{1}^{\mathrm{ad}}M_{1}(z|f,g)=  \notag \\
&&=[M_{1}(z|f),g]-(-1)^{\epsilon(f)\epsilon(g)}[M_{1}(z|g),f]-M_{1}(z|[f,g]).
\label{6.9}
\end{eqnarray}
Let
\[
[z\cup\mathrm{supp}(f)]\cap\mathrm{supp}(g)=\varnothing\;
\Longrightarrow\;[ M_1(z|g),f]=0\;\Longrightarrow\;\partial_A\hat{M}_1(z|g)=0
\;\Longrightarrow
\]
\begin{equation*}
M_{1}(z|f)=M_{1\mathrm{loc}}(z|f)+\int du\theta (x-u)t_{1|1}(u)f_{0}(u)+\int
du\theta (x-u)t_{1|2}(u)f_{1}(u)+t_{1|3}(f_{0})+t_{1|4}(f_{1}).
\end{equation*}
Let
\begin{equation*}
\lbrack \mathrm{supp}(f)\cup \mathrm{supp}(g)]>z\;\Longrightarrow
\end{equation*}
\begin{equation*}
c_{1}\int du[\partial
_{u}^{3}f_{1}(u)]g_{1}(u)=-t_{1|3}([f,g]_{0}+[g,f]_{0})-t_{1|4}([f,g]_{1})\;
\Longrightarrow
\end{equation*}
\begin{equation*}
c_{1}=t_{1|3}(f_{0})=t_{1|4}(f_{1})=0.
\end{equation*}
Let
\begin{equation*}
\lbrack \mathrm{supp}(f)\cup \mathrm{supp}(g)]<z\;\Longrightarrow
\end{equation*}
\begin{eqnarray*}
&&2c_{2}\int du[\partial _{u}^{3}f_{1}(u)]g_{1}(u)=
\int dut_{1|1}(u)\{[f(u),g(u)]_{0}+[g(u),f(u)]_{0}\}+ \\
&&+\int du\theta (x-u)t_{1|2}(u)[f(u),g(u)]_{1}\;\Longrightarrow\;
2c_{2}=t_{1|1}(u)=t_{1|2}(u)=0.
\end{eqnarray*}

Further treatment is common for arbitrary $n$.

\subsubsection{$n\geq 1$, $c_1=c_2=0$}

Let
\begin{eqnarray}
&&c_{3}m_{2|3}(z|f,g)+c_{4}m_{2|4}(z|f,g)=d_{1}^{\mathrm{ad}}M_{1}(z|f,g)=
\notag \\
&&\,=[M_{1}(z|f),g]-(-1)^{(\varepsilon (f)+1)(\varepsilon
(g)+1)}[M_{1}(z|g),f]-M_{1}(z|[f,g]).  \label{6.8}
\end{eqnarray}
Set
\begin{equation*}
z\bigcap \mathrm{supp}(f)=z\bigcap \mathrm{supp}(g)=\varnothing .
\end{equation*}
We have (according to Sec. \ref{H^{1}})
\[
\hat{M}_{1}(z|[f,,g])=0\;\Longrightarrow\;\;
M_1(z|f)=\sum_{q=0}^Qt^{(B)_q}(z)(\partial_B)^qf(z).
\]

Choosing $g(z)=1$, we obtain
\begin{equation*}
c_{3}f(z)-c_{3}N_\xi f(z)
+c_{4}\Delta f(z)=[t^{(B)_{0}}(z),f(z)]\;\Longrightarrow \;c_{3}=c_{4}=0.
\end{equation*}

So, we obtain that Eqs. (\ref{6.9}) and (\ref{6.8}) has solutions only if
$c_{1}=c_{2}=c_{3}=c_{4}=0$.

\subsection{The exactness of the form $M_{d|2}(z|f,g)$}

Let us discuss the terms $M_{d|2}(z|f,g)$ in Eq. (\ref{6.7a}),
\begin{equation*}
M_{d|2}(z|f,g)\equiv d_{1}^{\mathrm{ad}}M_{1}(z|f,g)=[M_{1}(z|f),g]-(-1)^{(
\varepsilon (f)+1)(\varepsilon (g)+1)}[M_{1}(z|g),f]-M_{1}(z|[f,g]).
\end{equation*}
Recall once again that the form $M_{d|2}(z|f,g)$ is exact if both functions $
M_{d|2}(z|f,g)$ and $M_{1}(z|g)$ in the expression $d_{1}^{\mathrm{ad}
}M_{1}(z|f,g)$ belong to the same space $\mathcal{A}$ for all $f,g\in \mathbf D_n$.

1) $\mathcal{A}=\mathbf D_n'$. In this case $M_{1}(z|f)\in \mathbf D_n'$ and
the form $M_{d|2}(z|f,g)$ is exact.

2) $\mathcal{A}=\mathbf E_n$. In this case $M_{1}(z|f)\in \mathbf E_n$
and
the form $M_{d|2}(z|f,g)$ is exact.\footnote{
\label{foot}
The proof can be found in \cite{SKT1}}

3) $\mathcal{A}=\mathbf D_n$. It is easy to prove that if
$M_{2}(z|f,g)\in\mathbf D_n$ then $c_1=c_2=0$
and $M_{d|2}(z|f,g)\in \mathbf D_n$.
Let $f(z)=\omega_{AB}z^B\tilde{f}(z)$, where $\tilde{f}(z)=1$ for
$z\in \mathrm{supp}(g)$. From $M_{d|2}(z|f,g)\in \mathbf D_n$, we obtain
\begin{equation}\label{42}
M_{1}(z|\partial_Ag)\in \mathbf D_n,\quad\forall A,g.
\end{equation}
Let $f(z)=[z^A(-1)^{\varepsilon_A}\omega_{AB}z^B/2n]\tilde{f}(z)$.
It follows from $M_{d|2}(z|f,g)\in \mathbf D_n$ and (\ref{42})
\[
M_{1}(z|[f,g])=M_{1}(z|\partial_Ag^{\prime A})+
M_{1}(z|g)\in \mathbf D_n,\quad g^{\prime
A}=\frac{1}{n}z^Ag(z),\quad \Longrightarrow
\]
\[
M_{1}(z|g)\in \mathbf D_n,\quad \forall g.
\]

Thus, the term $M_{d|2}(z|f,g)$ in Eq. (\ref{6.7a}) is the
exact form for any space $\mathcal{A}$ unlike
the case of the Poisson algebra.


\setcounter{MaxMatrixCols}{10}

\def\theequation{A\arabic{appen}.\arabic{equation}}

\renewcommand{\theorem}{\par\refstepcounter{theorem}
{\bf Theorem A\arabic{appen}.\arabic{theorem}. }}
\renewcommand{\lemma}{\par\refstepcounter{lemma}
{\bf Lemma A\arabic{appen}.\arabic{lemma}. }}
\renewcommand{\proposition}{\par\refstepcounter{proposition}
{\bf Proposition A\arabic{appen}.\arabic{proposition}. }}
\makeatletter \@addtoreset{theorem}{appen}
\makeatletter \@addtoreset{lemma}{appen}
\makeatletter \@addtoreset{proposition}{appen}
\makeatletter \@addtoreset{equation}{appen}
\renewcommand\thetheorem{A\theappen.\arabic{theorem}}
\renewcommand\thelemma{A\theappen.\arabic{lemma}}
\renewcommand\theproposition{A\theappen.\arabic{proposition}}

\appen{The proof of Proposition \ref{propadn=1}}\label{appproof}

Represent the forms $M_{1}(z|f)$ and $M_{2}(z|f,g)$ in the form
\begin{equation*}
M_{1}(z|f)=T_{(1)}(x|f_{0})+T_{(2)}(x|f_{1})+\xi \lbrack
T_{(3)}(x|f_{0})+T_{(4)}(x|f_{1})],
\end{equation*}
\begin{eqnarray*}
&&M_{2}(z|f,g)=M_{(1)}(x|f_{0},g_{0})+M_{(2)}(x|f_{0},g_{1})-M_{(2)}(x|g_{0},f_{1})+M_{(3)}(x|f_{1},g_{1})+
\\
&&+\xi \lbrack
M_{(4)}(x|f_{0},g_{0})+M_{(5)}(x|f_{0},g_{1})-M_{(5)}(x|g_{0},f_{1})+M_{(6)}(x|f_{1},g_{1})],
\\
&&M_{(1,4)}(x|\varphi ,\phi )=M_{(1,4)}(x|\phi ,\varphi
),\;M_{(3,6)}(x|\varphi ,\phi )=-M_{(3,6)}(x|\phi ,\varphi ).
\end{eqnarray*}
We have for $M_{2d}(z|f,g)\equiv d_{1}^{\mathrm{ad}}M_{1}(z|f,g)$:
\begin{eqnarray}
&&M_{d|(1)}(x|\varphi ,\phi )=-T_{(3)}(x|\varphi )\partial _{x}\phi
(x)-T_{(3)}(x|\phi )\partial _{x}\varphi (x),\;M_{d|(4)}(x|\varphi ,\phi )=0,
\label{5.2.1a} \\
&&M_{d|(2)}(x|\varphi ,\phi )=\partial _{x}T_{(1)}(x|\varphi )\phi
(x)+T_{(4)}(x|\phi )\partial _{x}\varphi (x)-T_{(1)}(x|[\varphi ,\phi ]_{0}),
\label{5.2.1b} \\
&&M_{d|(3)}(x|\varphi ,\phi )=\partial _{x}T_{(2)}(x|\varphi )\phi
(x)-\partial _{x}T_{(2)}(x|\phi )\varphi (x)-T_{(2)}(x|[\varphi ,\phi ]_{1}),
\label{5.2.1c} \\
&&M_{d|(5)}(x|\varphi ,\phi )=\partial _{x}T_{(3)}(x|\varphi )\phi
(x)-T_{(3)}(x|\varphi )\partial _{x}\phi (x)-T_{(3)}(x|[\varphi ,\phi ]_{0}),
\label{5.2.1d} \\
&&M_{d|(6)}(x|\varphi ,\phi )=\partial _{x}T_{(4)}(x|\varphi )\phi
(x)-\partial _{x}T_{(4)}(x|\phi )\varphi (x)+T_{(4)}(x|\phi )\partial
_{x}\varphi (x)-  \notag \\
&&-T_{(4)}(x|\varphi )\partial _{x}\phi (x)-T_{(4)}(x|[\varphi ,\phi ]_{1}),
\label{5.2.1e} \\
&&\,[\varphi (x),\phi (x)]_{0}=\{\partial _{x}\varphi (x)\}\phi
(x),\;[\varphi (x),\phi (x)]_{1}=\{\partial _{x}\varphi (x)\}\phi
(x)-\varphi (x)\partial _{x}\phi (x).  \notag
\end{eqnarray}

It follows from $d_{2}^{\mathrm{ad}}M_{2}(z|f_{0},g_{0},h_{0})=0$ and $
d_{2}^{\mathrm{ad}}M_{2}(z|f_{0},g_{0},h_{1})=0$
\begin{eqnarray}
&&M_{(4)}(x|\varphi ,\phi )\partial _{x}\omega (x)+\mathrm{cycle}(\varphi
,\phi ,\omega )=0,  \label{5.2.2a} \\
&&M_{(4)}(x|\varphi ,\phi )\partial _{x}\omega (x)-\{\partial
_{x}M_{(4)}(x|\varphi ,\phi )\}\omega (x)+M_{(4)}(x|[\varphi ,\omega
]_{0},\phi )+M_{(4)}(x|\varphi ,[\phi ,\omega ]_{0})=0,  \notag
\end{eqnarray}
\begin{eqnarray}
&&M_{(1)}(x|[\varphi ,\omega ]_{0},\phi )+M_{(1)}(x|\varphi ,[\phi ,\omega
]_{0})-\{\partial _{x}M_{(1)}(x|\varphi ,\phi )\}\omega (x)-  \notag \\
&&\,-M_{(5)}(x|\varphi ,\omega )\partial _{x}\phi (x)-M_{(5)}(x|\phi ,\omega
)\partial _{x}\varphi (x)=0.  \label{5.2.2c}
\end{eqnarray}

It follows from $d_{2}^{\mathrm{ad}}M_{2}(z|f_{0},g_{1},h_{1})=0$
\begin{eqnarray}
\!\!\!\!\!\!\!\!\!\!\!\!\!\!\!\!\!\!\!\!\!\!\!\!\!\!\!
&&M_{(5)}(x|\varphi ,\phi )\partial _{x}\omega (x)-\{\partial
_{x}M_{(5)}(x|\varphi ,\phi )\}\omega (x)+\{\partial _{x}M_{(5)}(x|\varphi
,\omega )\}\phi (x)-  \notag \\
\!\!\!\!\!\!\!\!\!\!\!\!\!\!\!\!\!\!\!\!\!\!\!\!\!\!\!
&&\,-M_{(5)}(x|\varphi ,\omega )\partial _{x}\phi (x)-M_{(5)}(x|[\varphi
,\phi ]_{0},\omega )+M_{(5)}(x|[\varphi ,\omega ]_{0},\phi
)+M_{(5)}(x|\varphi ,[\phi ,\omega ]_{1})=0,  \label{5.2.2b}
\\
\!\!\!\!\!\!\!\!\!\!\!\!\!\!\!\!\!\!\!\!\!\!\!\!\!\!\!
&&\ \ \ \ \ \ \ \{\partial _{x}M_{(2)}(x|\varphi ,\omega )\}\phi (x)-\{\partial
_{x}M_{(2)}(x|\varphi ,\phi )\}\omega (x)-M_{(2)}(x|[\varphi ,\phi
]_{0},\omega )+  \notag \\
\!\!\!\!\!\!\!\!\!\!\!\!\!\!\!\!\!\!\!\!\!\!\!\!\!\!\!
&&\ \ \ \ \ \ \ \,+M_{(2)}(x|[\varphi ,\omega ]_{0},\phi )+M_{(2)}(x|\varphi ,[\phi
,\omega ]_{1})+M_{(6)}(x|\phi ,\omega )\partial _{x}\varphi (x)=0.
\label{5.2.2e}
\end{eqnarray}

It follows from $d_{2}^{\mathrm{ad}}M_{2}(z|f_{1},g_{1},h_{1})=0$
\begin{eqnarray}
&&M_{(6)}(x|\varphi ,\phi )\partial _{x}\omega (x)-\{\partial
_{x}M_{(6)}(x|\varphi ,\phi )\}\omega (x)-M_{(6)}(x|[\varphi ,\phi
]_{1},\omega )+  \notag \\
&&\,+\mathrm{cycle}(\varphi ,\phi ,\omega )=0,  \label{5.2.2d}
\\
&&\ \ \ \ \ \ \ \
-[\{\partial _{x}M_{(3)}(x|\varphi ,\phi )\}\omega (x)+M_{(3)}(x|[\varphi
,\phi ]_{1},\omega )+  \notag \\
&&\ \ \ \ \ \ \ \ \,+\mathrm{cycle}(\varphi ,\phi ,\omega )]\equiv d_{2}^{\mathrm{ad}
}M_{(3)}(x|\varphi ,\phi ,\omega )=0.  \label{5.2.2f}
\end{eqnarray}

I. Consider Eq. (\ref{5.2.2a}). Let $x\subset U$, $U$ is some fixed bounded
domain, $\psi (x)\subset D$ is some fixed function, $\psi (x)=x$ for $
x\subset U$. We have from Eq. (\ref{5.2.2a})
\begin{equation*}
M_{(4)}(x|\varphi ,\phi )=\mu (x|\varphi )\partial _{x}\phi (x)+\mu (x|\phi
)\partial _{x}\varphi (x),\;\mu (x|\varphi )=-M_{(4)}(x|\varphi ,\psi
)\;\Longrightarrow
\end{equation*}
\begin{equation*}
\mu (x|\varphi )\partial _{x}\phi (x)\partial _{x}\omega (x)+\mu (x|\phi
)\partial _{x}\varphi (x)\partial _{x}\omega (x)+\mu (x|\omega )\partial
_{x}\varphi (x)\partial _{x}\phi (x)=0\;\Longrightarrow
\end{equation*}
\begin{equation*}
\mu (x|\varphi )=\mu (x)\partial _{x}\varphi (x),\;\mu (x)=-2\mu (x|\psi
)\;\Longrightarrow
\end{equation*}
\begin{equation*}
\mu (x)\partial _{x}\varphi (x)\partial _{x}\phi (x)\partial _{x}\omega
(x)=0\;\Longrightarrow \;\mu (x)=0\;\Longrightarrow \;\mu (x|\varphi
)=0\;\Longrightarrow
\end{equation*}
\begin{equation*}
M_{(4)}(x|\varphi ,\phi )=0.
\end{equation*}

II. Consider Eq. (\ref{5.2.2b}).

Let
$
\lbrack x\cup \mathrm{supp}(\varphi )]\cap \lbrack \mathrm{supp}(\phi )\cup
\mathrm{supp}(\omega )]=\varnothing .
$

We have
$\hat{M}_{(5)}(x|\varphi ,[\phi ,\omega ]_{1})=0\;\Longrightarrow \;\hat{M}
_{(5)}(x|\varphi ,\phi )=0\;\Longrightarrow
$
\begin{eqnarray*}
M_{(5)}(x|\varphi ,\phi ) &=&M_{(5)1}(x|\varphi ,\phi )+M_{(5)2}(x|\varphi
,\phi ), \\
M_{(5)1}(x|\varphi ,\phi ) &=&\sum_{q=0}^{Q}M_{2}^{q}(x|\varphi )\partial
_{x}^{q}\phi (x),\;M_{(5)2}(x|\varphi ,\phi
)=\sum_{q=0}^{Q}M_{2}^{q}(x|\{\partial ^{q}\varphi \}\phi ).
\end{eqnarray*}

Let
$\lbrack x\cup \mathrm{supp}(\omega )]\cap \lbrack \mathrm{supp}(\varphi
)\cup \mathrm{supp}(\phi )]=\varnothing .
$

We have
\begin{align*}
\sum_{q=0}^{Q}\hat{M}_{2}^{q}(x|\{\partial ^{q}\varphi \}\phi )\partial
_{x}\omega (x)-\sum_{q=0}^{Q}\partial _{x}\hat{M}_{2}^{q}(x|\{\partial
^{q}\varphi \}\phi )\omega (x)-
\sum_{q=0}^{Q}\hat{M}_{1}^{q}(x|[\varphi ,\phi ]_{0})\partial
_{x}^{q}\omega (x)=0\;\Longrightarrow
\end{align*}
\begin{eqnarray*}
\hat{M}_{1}^{0}(x|[\varphi ,\phi ]_{0}) =-\sum_{q=0}^{Q}\partial _{x}\hat{M
}_{2}^{q}(x|\{\partial ^{q}\varphi \}\phi ), \quad
&&
\hat{M}_{1}^{1}(x|[\varphi ,\phi ]_{0}) =\sum_{q=0}^{Q}\hat{M}
_{2}^{q}(x|\{\partial ^{q}\varphi \}\phi ); \\
&&
\hat{M}_{1}^{q}(x|\varphi ) =0,\;\forall q\geq 2\;\Longrightarrow
\end{eqnarray*}
\begin{align*}
& M_{(5)2}(x|\varphi ,\phi )=-T_{(3)}(x|[\varphi ,\phi ]_{0})+M_{(5)2\mathrm{
loc}}(x|\varphi ,\phi ), \\
& M_{(5)1}(x|\varphi ,\phi )=\partial _{x}T_{(3)}(x|\varphi )\phi
(x)-T_{(3)}(x|\varphi )\partial _{x}\phi (x)+M_{(5)1\mathrm{loc}}(x|\varphi
,\phi ), \\
& \hat{M}_{2}^{q}(x|\varphi )=0,\;q\neq 1,\;T_{(3)}(x|\varphi
)=-M_{2}^{1}(x|\varphi ).
\end{align*}
So, we obtain
\begin{equation*}
M_{(5)}(x|\varphi ,\phi )=M_{(5)\mathrm{loc}}(x|\varphi ,\phi
)+M_{d|(5)}(x|\varphi ,\phi ),
\end{equation*}
where the expression for $M_{d|(5)}(x|\varphi ,\phi )$ is given by Eq. (\ref
{5.2.1d}).

III. Consider Eq. (\ref{5.2.2c}).

Let
$\lbrack x\cup \mathrm{supp}(\varphi )]\cap \lbrack \mathrm{supp}(\phi )\cup
\mathrm{supp}(\omega )]=\varnothing .
$

We have for $M_{(1)}^{\prime }(x|\varphi ,\phi )=M_{(1)}(x|\varphi ,\phi
)-M_{d|(1)}(x|\varphi ,\phi )$, the expression for $M_{d|(5)}(x|\varphi
,\phi )$ is given by Eq. (\ref{5.2.1a}),
\begin{align*}
& \hat{M}_{(1)}^{\prime }(x|\varphi ,[\phi ,\omega ]_{0})=0\;\Longrightarrow
\\
& M_{(1)}^{\prime }(x|\varphi ,\phi )=\sum_{q=0}^{Q}M_{3}^{q}(x|\varphi
)\partial _{x}^{q}\phi (x)=\sum_{q=0}^{Q}M_{3}^{q}(x|\phi )\partial
_{x}^{q}\varphi (x)\;\Longrightarrow
\end{align*}
\begin{equation*}
M_{(1)}^{\prime }(x|\varphi ,\phi )=M_{(1)\mathrm{loc}}(x|\varphi ,\phi
),\;M_{(1)}(x|\varphi ,\phi )=M_{(1)\mathrm{loc}}(x|\varphi ,\phi
)+M_{d|(1)}(x|\varphi ,\phi ).
\end{equation*}

IV. Consider Eq. (\ref{5.2.2d}).

Let
$\lbrack x\cup \mathrm{supp}(\omega )]\cap \lbrack \mathrm{supp}(\varphi
)\cup \mathrm{supp}(\phi )]=\mathrm{supp}(\varphi )\cap \mathrm{supp}(\phi
)=\varnothing .
$

We have
$\hat{M}_{(6)}(x|\varphi ,\phi )\partial _{x}\omega (x)-\{\partial _{x}\hat{M}
_{(6)}(x|\varphi ,\phi )\}\omega (x)=0
$ and so
$\hat{M}_{(6)}(x|\varphi ,\phi )=0.
$
Thus
\begin{eqnarray*}
M_{(6)}(x|\varphi ,\phi ) &=&M_{(6)4}(x|\varphi ,\phi )+M_{(6)5}(x|\varphi
,\phi ), \\
M_{(6)4}(x|\varphi ,\phi ) &=&\sum_{q=0}^{Q}\{\partial _{x}^{q}\varphi
(x)M_{4}^{q}(x|\phi )-M_{4}^{q}(x|\varphi )\partial _{x}^{q}\phi (x)\}, \\
M_{(6)5}(x|\varphi ,\phi ) &=&\sum_{k=0}^{K}M_{5}^{2k+1}(x|\{\partial
^{2k+1}\varphi \}\phi -\varphi \partial ^{2k+1}\phi ).
\end{eqnarray*}

Let
$\lbrack x\cup \mathrm{supp}(\omega )]\cap \lbrack \mathrm{supp}(\varphi
)\cup \mathrm{supp}(\phi )]=\varnothing .
$

We have
\begin{align*}
& \sum_{k=0}^{K}\hat{M}_{5}^{2k+1}(x|\{\partial ^{2k+1}\varphi \}\phi
-\varphi \partial ^{2k+1}\phi )\partial _{x}\omega
(x)-\sum_{k=0}^{K}\partial _{x}\hat{M}_{5}^{2k+1}(x|\{\partial
^{2k+1}\varphi \}\phi -\varphi \partial ^{2k+1}\phi )\omega (x)- \\
& \,+\sum_{q=0}^{Q}\hat{M}_{4}^{q}(x|\{\partial \varphi \}\phi -\varphi
\partial \phi )\partial _{x}^{q}\omega (x)=0.
\end{align*}
So
\begin{eqnarray*}
\hat{M}_{4}^{0}(x|\{\partial \varphi \}\phi -\varphi \partial \phi )
&=&\sum_{k=0}^{K}\partial _{x}\hat{M}_{5}^{2k+1}(x|\{\partial ^{2k+1}\varphi
\}\phi -\varphi \partial ^{2k+1}\phi ), \\
\hat{M}_{4}^{1}(x|\{\partial \varphi \}\phi -\varphi \partial \phi )
&=&-\sum_{k=0}^{K}\hat{M}_{5}^{2k+1}(x|\{\partial ^{2k+1}\varphi \}\phi
-\varphi \partial ^{2k+1}\phi ), \\
\hat{M}_{4}^{q}(x|\varphi ) &=&0\;\forall q\geq 2.\;\Longrightarrow
\end{eqnarray*}
\begin{eqnarray*}
&&M_{(6)4}(x|\varphi ,\phi )=\partial _{x}T_{(4)}(x|\phi )\varphi
(x)-\partial _{x}T_{(4)}(x|\varphi )\phi (x)+ \\
&&+T_{(4)}(x|\phi )\partial _{x}\varphi (x)-T_{(4)}(x|\varphi )\partial
_{x}\phi (x)+M_{(6)4\mathrm{lok}}(x|\varphi ,\phi ), \\
&&M_{(6)5}(x|\varphi ,\phi )=-T_{(4)}(x|[\varphi ,\phi ]_{1})+M_{(6)5\mathrm{
lok}}(x|\varphi ,\phi ), \\
&&\hat{M}_{5}^{2k+1}(x|\varphi )=0,\;k\geq 1,\;T_{(4)}(x|\varphi
)=-M_{5}^{1}(x|\varphi ).
\end{eqnarray*}
So, we obtain
\begin{equation*}
M_{(6)}(x|\varphi ,\phi )=M_{(6)\mathrm{loc}}(x|\varphi ,\phi
)+M_{d|(6)}(x|\varphi ,\phi ),
\end{equation*}
where the expression for $M_{d|(6)}(x|\varphi ,\phi )$ is given by Eq. (\ref
{5.2.1e}).

V. Consider Eq. (\ref{5.2.2e}).

We have for $M_{(2)}^{\prime }(x|\varphi ,\phi )=M_{(2)}(x|\varphi ,\phi
)-T_{(4)}(x|\phi )\partial _{x}\varphi (x)$
\begin{eqnarray*}
&&\{\partial _{x}M_{(2)}^{\prime }(x|\varphi ,\omega )\}\phi (x)-\{\partial
_{x}M_{(2)}^{\prime }(x|\varphi ,\phi )\}\omega (x)-M_{(2)}^{\prime
}(x|[\varphi ,\phi ]_{0},\omega )+ \\
&&\,+M_{(2)}^{\prime }(x|[\varphi ,\omega ]_{0},\phi )+M_{(2)}^{\prime
}(x|\varphi ,[\phi ,\omega ]_{1})+M_{(6)\mathrm{loc}}(x|\phi ,\omega
)\partial _{x}\varphi (x)=0.
\end{eqnarray*}

Let
$\lbrack x\cup \mathrm{supp}(\varphi )]\cap \lbrack \mathrm{supp}(\phi )\cup
\mathrm{supp}(\omega )]=\varnothing .
$

We have
$\hat{M}_{(2)}^{\prime }(x|\varphi ,[\phi ,\omega ]_{1})=0$. So
$\hat{M}_{(2)}^{\prime }(x|\varphi ,\phi )=0$
and
\begin{eqnarray*}
M_{(2)}^{\prime }(x|\varphi ,\phi ) &=&M_{(2)6}(x|\varphi ,\phi
)+M_{(2)7}(x|\varphi ,\phi ), \\
M_{(2)6}(x|\varphi ,\phi ) &=&\sum_{q=0}^{Q}M_{6}^{q}(x|\varphi )\partial
_{x}^{q}\phi (x),\;M_{(2)7}(x|\varphi ,\phi
)=\sum_{q=0}^{Q}M_{7}^{q}(x|\{\partial ^{q}\varphi \}\phi ).
\end{eqnarray*}

Let
$\lbrack x\cup \mathrm{supp}(\omega )]\cap \lbrack \mathrm{supp}(\varphi
)\cup \mathrm{supp}(\phi )]=\varnothing .
$

We obtain
$\{\partial _{x}\hat{M}_{(2)}^{\prime }(x|\varphi ,\phi )\}\omega (x)+\hat{M}
_{(2)}^{\prime }(x|[\varphi ,\phi ]_{0},\omega )=0
$
and so
\begin{equation*}
\sum_{q=0}^{Q}\{\partial _{x}\hat{M}_{7}^{q}(x|\{\partial ^{q}\varphi \}\phi
)\}\omega (x)+\sum_{q=0}^{Q}\hat{M}_{6}^{q}(x|\{\partial \varphi \}\phi
)\partial _{x}^{q}\omega (x)=0\;\Longrightarrow
\end{equation*}
\begin{equation*}
\sum_{q=0}^{Q}\partial _{x}\hat{M}_{7}^{q}(x|\{\partial ^{q}\varphi \}\phi )+
\hat{M}_{6}^{0}(x|\{\partial \varphi \}\phi )=0,\;\hat{M}_{6}^{q}(x|\varphi
)=0,\;q\geq 1\;\Longrightarrow
\end{equation*}
\begin{equation*}
\hat{M}_{6}^{0}(x|\varphi )=-\partial _{x}\hat{M}_{7}^{1}(x|\varphi
),\;\partial _{x}\hat{M}_{7}^{q}(x|\varphi )=0,\;q\neq 1
\end{equation*}
So, we found
\begin{eqnarray*}
M_{(2)}(x|\varphi ,\phi ) &=&M_{(2)\mathrm{loc}}(x|\varphi ,\phi
)+M_{d|(2)}(x|\varphi ,\phi )+M_{(2)}^{\prime \prime }(x|\varphi ,\phi ), \\
M_{(2)}^{\prime \prime }(x|\varphi ,\phi ) &=&\sum_{q=0,q\neq
1}^{Q}M_{7}^{q}(x|\{\partial ^{q}\varphi \}\phi ),\;\partial _{x}\hat{M}
_{7}^{q}(x|\varphi )=0,
\end{eqnarray*}
where the expression for $M_{d|(2)}(x|\varphi ,\phi )$ is given by Eq. (\ref
{5.2.1b}) with $T_{(1)}(x|\phi )=-M_{7}^{1}(x|\varphi )$. For $
M_{(2)}^{\prime \prime }(x|\varphi ,\phi )$ we obtain
\begin{eqnarray*}
&&\{\partial _{x}M_{(2)}^{\prime \prime }(x|\varphi ,\omega )\}\phi
(x)-\{\partial _{x}M_{(2)}^{\prime \prime }(x|\varphi ,\phi )\}\omega
(x)-M_{(2)}^{\prime \prime }(x|[\varphi ,\phi ]_{0},\omega )+ \\
&&\,+M_{(2)}^{\prime \prime }(x|[\varphi ,\omega ]_{0},\phi
)+M_{(2)}^{\prime \prime }(x|\varphi ,[\phi ,\omega ]_{1})=M_{\mathrm{loc}
}(x|\varphi ,\phi ,\omega ),
\end{eqnarray*}
where $M_{\mathrm{loc}}(x|\varphi ,\phi ,\omega )$ is some local over all
arguments functional.

Let
$x\cap \lbrack \mathrm{supp}(\varphi )\cup \mathrm{supp}(\phi )\cup \mathrm{
supp}(\omega )]=\varnothing .
$

We obtain
\begin{equation*}
\hat{M}_{(2)}^{\prime \prime }(x|[\varphi ,\phi ]_{0},\omega )-\hat{M}
_{(2)}^{\prime \prime }(x|[\varphi ,\omega ]_{0},\phi )-\hat{M}
_{(2)}^{\prime \prime }(x|\varphi ,[\phi ,\omega ]_{1})=0
\end{equation*}
or
\begin{equation}
\sum_{q=0,q\neq 1}^{Q}\hat{M}_{7}^{q}(x|\{\partial ^{q}(\partial \varphi
\phi )\}\omega -\{\partial ^{q}(\partial \varphi \omega )\}\phi -\{\partial
^{q}\varphi \}[\partial \phi \omega -\phi \partial \omega ])=0.
\label{5.2.3}
\end{equation}
Let $\varphi (x)\rightarrow e^{px}\varphi (x)$, $\phi (x)\rightarrow
e^{kx}\phi (x)$, $\omega (x)\rightarrow e^{-(p+k)x}\omega (x)$. Consider the
terms of highest order in $p$, $k$ in Eq. (\ref{5.2.3}),
\begin{equation*}
\lbrack p(p+k)^{Q}-p(-k)^{Q}-(p+2k)p^{Q}]\hat{M}_{7}^{Q}(x|\varphi \phi
\omega )=0\;\Longrightarrow
\end{equation*}
\begin{equation*}
\hat{M}_{7}^{Q}(x|\varphi )=0,\;Q\neq 1.
\end{equation*}
Finally, we have
\begin{equation*}
M_{(2)}(x|\varphi ,\phi )=M_{(2)\mathrm{loc}}(x|\varphi ,\phi
)+M_{d|(2)}(x|\varphi ,\phi ).
\end{equation*}

VI. Consider Eq. (\ref{5.2.2f}).

Let
$\lbrack x\cup \mathrm{supp}(\omega )]\cap \lbrack \mathrm{supp}(\varphi
)\cup \mathrm{supp}(\phi )]=x\cap \mathrm{supp}(\omega )=\varnothing .
$

We have
\begin{equation*}
\hat{M}_{(3)}(x|[\varphi ,\phi ]_{1},\omega )=0\;\Longrightarrow \;\hat{M}
_{(3)}(x|\varphi ,\omega )=0\;\Longrightarrow
\end{equation*}
\begin{eqnarray*}
M_{(3)}(x|\varphi ,\phi ) &=&M_{(3)8}(x|\varphi ,\phi )+M_{(3)9}(x|\varphi
,\phi ), \\
M_{(3)8}(x|\varphi ,\phi ) &=&\sum_{q=0}^{Q}\{\partial _{x}^{q}\varphi
(x)M_{8}^{q}(x|\phi )-M_{8}^{q}(x|\varphi )\partial _{x}^{q}\phi (x)\}, \\
M_{(3)9}(x|\varphi ,\phi ) &=&\sum_{l=0}^{L}M_{9}^{2l+1}(x|\{\partial
^{2l+1}\varphi \}\phi -\varphi \partial ^{2l+1}\phi ).
\end{eqnarray*}

Let
$\lbrack x\cup \mathrm{supp}(\omega )]\cap \lbrack \mathrm{supp}(\varphi
)\cup \mathrm{supp}(\phi )]=\varnothing .
$

We have
$\{\partial _{x}\hat{M}_{(3)}(x|\varphi ,\phi )\}\omega (x)+\hat{M}
_{(3)}(x|[\varphi ,\phi ]_{1},\omega )=0
$
or
$\sum_{l=0}^{L}[\partial _{x}\hat{M}_{9}^{2l+1}(x|\{\partial ^{2l+1}\varphi
\}\phi -\varphi \partial ^{2l+1}\phi )]\omega (x)-\sum_{q=0}^{Q}\hat{M}
_{8}^{q}(x|\partial \varphi \phi -\varphi \partial \phi )\partial
_{x}^{q}\omega (x)\}=0$.
So
\begin{eqnarray*}
&&\sum_{l=0}^{L}[\partial _{x}\hat{M}_{9}^{2l+1}(x|\{\partial ^{2l+1}\varphi
\}\phi -\varphi \partial ^{2l+1}\phi )]-\hat{M}_{8}^{0}(x|\partial \varphi
\phi -\varphi \partial \phi )=0, \\
&&\hat{M}_{8}^{q}(x|\varphi )=0,\;q\geq 1\;\Longrightarrow
\end{eqnarray*}
\begin{equation*}
\hat{M}_{8}^{0}(x|\varphi )=\partial _{x}\hat{M}_{9}^{1}(x|\varphi
),\;\partial _{x}\hat{M}_{9}^{2l+1}(x|\varphi )=0,\;l\geq 1.
\end{equation*}
Thus, we found
\begin{eqnarray*}
M_{(3)}(x|\varphi ,\phi ) &=&M_{d|(3)}(x|\varphi ,\phi )+M_{(3)}^{\prime
}(x|\varphi ,\phi )+M_{(3)\mathrm{lok}}(x|\varphi ,\phi ), \\
M_{(3)}^{\prime }(x|\varphi ,\phi )
&=&\sum_{l=1}^{L}M_{9}^{2l+1}(x|\{\partial ^{2l+1}\varphi \}\phi -\varphi
\partial ^{2l+1}\phi ),
\end{eqnarray*}
where the expression for $M_{d|(3)}(x|\varphi ,\phi )$ is given by Eq. (\ref
{5.2.1c}) with $T_{(2)}(x|\phi )=-M_{9}^{1}(x|\varphi )$. For $
M_{(3)}^{\prime }(x|\varphi ,\phi )$ we obtain
\begin{eqnarray*}
&&\{\partial _{x}M_{(3)}^{\prime }(x|\varphi ,\omega )\}\phi (x)-\{\partial
_{x}M_{(3)}^{\prime }(x|\varphi ,\phi )\}\omega (x)-\{\partial
_{x}M_{(3)}^{\prime }(x|\phi ,\omega )\}\varphi (x)- \\
&&\,-M_{(3)}^{\prime }(x|[\varphi ,\phi ]_{1},\omega )+M_{(3)}^{\prime
}(x|[\varphi ,\omega ]_{1},\phi )+M_{(3)}^{\prime }(x|\varphi ,[\phi ,\omega
]_{1})=M_{\mathrm{loc}}^{\prime }(x|\varphi ,\phi ,\omega ).
\end{eqnarray*}
where $M_{\mathrm{loc}}^{\prime }(x|\varphi ,\phi ,\omega )$ is some local
over all arguments functional.

Let
$
x\cap \lbrack \mathrm{supp}(\varphi )\cup \mathrm{supp}(\phi )\cup \mathrm{
supp}(\omega )]=\varnothing .
$

We obtain
$\hat{M}_{(3)}^{\prime }(x|[\varphi ,\phi ]_{1},\omega )+\hat{M}
_{(3)}^{\prime }(x|[\omega ,\varphi ]_{1},\phi )+\hat{M}_{(3)}^{\prime
}(x|[\phi ,\omega ]_{1},\varphi )=0
$
or
\begin{equation}
\sum_{l=1}^{L}\hat{M}_{9}^{2l+1}(x|\{\partial ^{2l+1}[\varphi ,\phi
]_{1}\}\omega -\{\partial ^{2l+1}\omega \}[\varphi ,\phi ]_{1}+\mathrm{cycle}
(\varphi ,\phi ,\omega ))=0.  \label{5.2.4}
\end{equation}
Let $\varphi (x)\rightarrow e^{px}\varphi (x)$, $\phi (x)\rightarrow
e^{kx}\phi (x)$, $\omega (x)\rightarrow e^{-(p+k)x}\omega (x)$. Consider the
terms of highest order in $p$, $k$ in Eq. (\ref{5.2.4}),
\begin{equation*}
\lbrack (p-k)(p+k)^{2L+1}+(2p+k)k^{2L+1}-(p+2k)p^{2L+1}]\hat{M}
_{9}^{2L+1}(x|\varphi \phi \omega )=0\;\Longrightarrow
\end{equation*}
\begin{equation*}
\hat{M}_{9}^{2L+1}(x|\varphi )=0,\;L\geq 2\;\Longrightarrow
\end{equation*}
\begin{eqnarray}
&&M_{(3)}^{\prime }(x|\varphi ,\phi )=M_{9}^{3}(x|\{\partial ^{3}\varphi
\}\phi -\varphi \partial ^{3}\phi ),  \notag \\
&&\partial _{x}\hat{M}_{9}^{3}(x|\varphi )=0,\;\partial _{x}\hat{m}
_{9}^{3}(x|u)=0,  \label{5.2.5b} \\
&&\hat{M}_{9}^{3}(x|\{\partial ^{3}[\varphi ,\phi ]_{1}\}\omega -\{\partial
^{3}\omega \}[\varphi ,\phi ]_{1}+\mathrm{cycle}(\varphi ,\phi ,\omega ))=0,
\label{5.2.5c} \\
&&M_{9}^{3}(x|\varphi )=\int dum_{9}^{3}(x|u)\varphi (u).  \notag
\end{eqnarray}
Let $\omega (x)=1$ in Eq. (\ref{5.2.5c}),
\begin{equation}
\hat{M}_{9}^{3}(x|[2\overleftarrow{\partial }\partial ^{3}\varphi +3
\overleftarrow{\partial }^{2}\partial ^{2}\varphi +\overleftarrow{\partial }
^{3}\partial \varphi ]\phi )=0\;\Longrightarrow \;\hat{m}_{9}^{3}(x|u)
\overleftarrow{\partial _{u}}=0.  \label{5.2.6}
\end{equation}

It follows from Eq. (\ref{5.2.5b})
\begin{eqnarray*}
&&\partial _{x}m_{9}^{3}(x|u)=\partial _{x}\sum_{q=0}m_{9}^{3q}(x)\partial
_{x}^{q}\delta (x-u)+\delta (x-u)\mu _{91}^{3}(u)= \\
&&\,=\partial _{x}\left( \sum_{q=0}m_{9}^{3q}(x)\partial _{x}^{q}\delta
(x-u)+\theta (x-u)\mu _{91}^{3}(u)\right) \;\Longrightarrow
\end{eqnarray*}
\begin{equation*}
m_{9}^{3}(x|u)=\sum_{q=0}m_{9}^{3q}(x)\partial _{x}^{q}\delta (x-u)+\theta
(x-u)\mu _{91}^{3}(u)+\mu _{92}^{3}(u).
\end{equation*}

It follows from Eq. (\ref{5.2.6})
\begin{equation*}
\theta (x-u)\partial _{u}\mu _{91}^{3}(u)+\partial _{u}\mu
_{92}^{3}(u)=0,\;x\neq u\;\Longrightarrow
\end{equation*}
\begin{equation*}
\mu _{91}^{3}(u)=c_{2}=\mathrm{const},\;\mu _{92}^{3}(u)=\frac{1}{2}c_{1}=
\mathrm{const}.
\end{equation*}
So, we obtain
\begin{eqnarray*}
M_{(3)}(x|\varphi ,\phi ) &=&c_{1}M_{(3)1}(x|\varphi ,\phi
)+M_{d|(3)}(x|\varphi ,\phi )+M_{(3)}^{\prime \prime }(x|\varphi ,\phi
)+M_{(3)\mathrm{lok}}(x|\varphi ,\phi ), \\
M_{(3)1}(x|\varphi ,\phi ) &=&\int du[\partial _{u}^{3}\varphi (u)]\phi (u),
\\
M_{(3)}^{\prime \prime }(x|\varphi ,\phi ) &=&c_{2}\int du\theta
(x-u)[\{\partial _{u}^{3}\varphi (u)\}\phi (u)-\varphi (u)\partial
_{u}^{3}\phi (u)].
\end{eqnarray*}
\begin{eqnarray*}
&&d_{2}^{\mathrm{ad}}M_{(3)}^{\prime \prime }(x|\varphi ,\phi ,\omega )=-\mu
_{91}^{3}[\{\partial _{x}^{2}\varphi (x)\}\phi (x)\partial _{x}\omega
(x)-\{\partial _{x}^{2}\varphi (x)\}\{\partial _{x}\phi (x)\}\omega (x)]+ \\
&&+\mathrm{cycle}(\varphi ,\phi ,\omega ).
\end{eqnarray*}
Note that the form
\begin{equation}
m_{2|1}(x|f,g)=M_{(3)1}(x|f_{1},g_{1})=\int du[\partial
^{3}f_{1}(u)]g_{1}(u),\;\varepsilon _{m_{2|1}}=0,  \label{5.2.7a}
\end{equation}
satisfies the cohomology equation (\ref{5.1}). Furthermore, a form $
c_{2}m_{2|2}(x|f,g),$ which differs from $M_{(3)}^{\prime \prime
}(x|f_{1},g_{1})$ by local form, satisfies the cohomology equation (\ref
{5.1}) too,
\begin{eqnarray}
&&m_{2|2}(x|f,g)=\int du\theta (x-u)[\{\partial
_{u}^{3}f_{1}(u)\}g_{1}(u)-f_{1}(u)\partial _{u}^{3}g_{1}(u)]+  \notag \\
&&+x[\{\partial _{x}^{2}f_{1}(x)\}\partial _{x}g_{1}(x)-\{\partial
_{x}f_{1}(x)\}\partial _{x}^{2}g_{1}(x)],\;\varepsilon _{m_{2|2}}=0.
\label{5.2.8a}
\end{eqnarray}

So, we obtained
\begin{equation*}
M_{2}(z|f,g)=c_{1}m_{2|1}(x|f,g)+c_{2}m_{2|2}(x|f,g)+d_{1}^{\mathrm{ad}
}M_{1}(z|f,g)+M_{2\mathrm{loc}}(z|f,g).
\end{equation*}

\appen{Lie Superalgebra PE}

Let $
p_{i}=x_i,\ p_{n+\alpha}=\xi _{\alpha},\ q_{i}=y_i,\ q_{n+\alpha}=\eta _{\alpha},\
\ i,j,...,\alpha ,\beta
,...=1,2,...,n
$.

Let
$
\overleftarrow{L}^{AB} =\overleftarrow{\partial }_{A}p_{C}\omega
^{CB}+(-1)^{\varepsilon _{A}\varepsilon _{B}}\overleftarrow{\partial }
_{B}p_{C}\omega ^{CA}$,
$\varepsilon \left( \overleftarrow{L}^{AB}\right) =\varepsilon
_{A}+\varepsilon _{B}+1$.
Introduce as well the notation
\begin{eqnarray*}
\overleftarrow{M}_{ij} &=&-\overleftarrow{L}^{ij}=\frac{\overleftarrow{
\partial }}{\partial x_{i}}\xi _{j}+\frac{\overleftarrow{\partial }}{
\partial x_{j}}\xi _{i}, \\
\;\overleftarrow{P}_{i\alpha } &=&\overleftarrow{L}^{i,n+\alpha }=\frac{
\overleftarrow{\partial }}{\partial x_{i}}x_{\alpha }-\frac{\overleftarrow{
\partial }}{\partial \xi _{\alpha }}\xi _{i}, \\
\overleftarrow{Q}_{\alpha \beta } &=&\overleftarrow{L}^{n+\alpha ,n+\beta }=
\frac{\overleftarrow{\partial }}{\partial \xi _{\alpha }}x_{\beta }-\frac{
\overleftarrow{\partial }}{\partial \xi _{\beta }}x_{\alpha },
\end{eqnarray*}
\begin{equation*}
\overleftarrow{\Sigma }_{p}=\overleftarrow{P}_{p\,ii}=\overleftarrow{N}_{x}-
\overleftarrow{N}_{\xi },\;\overleftarrow{N}_{x}=\frac{\overleftarrow{
\partial }}{\partial x_{i}}x_{i},\;\overleftarrow{N}_{\xi }=\frac{
\overleftarrow{\partial }}{\partial \xi _{i}}\xi _{i}.
\end{equation*}

These operators form the periplectic superalgebra ${\fr pe}(n)$ \cite{pe}, \cite{Leites}:
\begin{eqnarray*}
&&\left[ \overleftarrow{L}^{AB},\overleftarrow{L}^{CD}\right] = \\
&&-\left( \omega ^{BC}\overleftarrow{L}^{AD}+(-1)^{\varepsilon
_{C}\varepsilon _{D}}\omega ^{BD}\overleftarrow{L}^{AC}+(-1)^{\varepsilon
_{A}\varepsilon _{B}}\omega ^{AC}\overleftarrow{L}^{BD}+(-1)^{\varepsilon
_{A}\varepsilon _{B}+\varepsilon _{C}\varepsilon _{D}}\omega ^{AD}
\overleftarrow{L}^{BC}\right) ,
\end{eqnarray*}
and $\overleftarrow{\Sigma }_{p}$ generates $U(1)$ center of the even subalgebra
of this algebra.
Algebra ${\fr spe}(n)={\fr pe}(n)/U(1)$
is also known as strange Lie superalgebra $P(n-1)$ \cite{kac}.

We have
$
z_{A}(\overleftarrow{L}_{p}^{BC}+\overleftarrow{L}_{q}^{BC}) =z_{A}
\overleftarrow{L}_{z}^{BC},\;z_{A}=p_{A}+q_{A}$,
$\left\langle p,p\right\rangle \overleftarrow{L}_{p}^{AB} =\left\langle
p,q\right\rangle (\overleftarrow{L}_{p}^{BC}+\overleftarrow{L}_{q}^{BC})=0$,
where
$$
\left\langle p,q\right\rangle =p_{A}(-1)^{\varepsilon _{A}}\omega
^{AB}q_{B}=\left\langle q,p\right\rangle .
$$

\appen{The proof of Proposition \ref{eqSPE}}\label{SPE}

Here and in successive appendices we solve the equation

\begin{eqnarray}
&&F^{AB}(p)\overleftarrow{L}_{p}^{CD}-(-1)^{(\varepsilon _{C}+\varepsilon
_{D}+1)(\varepsilon _{A}+\varepsilon _{B}+1)}F^{CD}(p)\overleftarrow{L}
_{p}^{AB}=-\{F^{AD}(p)\omega ^{BC}+  \notag \\
&&\,+(-1)^{\varepsilon _{C}\varepsilon _{D}}F^{AC}(p)\omega
^{BD}+(-1)^{\varepsilon _{A}\varepsilon _{B}}F^{BD}(p)\omega
^{AC}+(-1)^{\varepsilon _{A}\varepsilon _{B}+\varepsilon _{C}\varepsilon
_{D}}F^{BC}(p)\omega ^{AD}\},  \label{3.1_}
\end{eqnarray}
or
\begin{eqnarray}
&&U^{ij}\overleftarrow{M}_{kl}+U^{kl}\overleftarrow{M}_{ij}=0,  \label{3.2_}
\\
&&U^{ij}\overleftarrow{P}_{k\alpha }+U^{ik}\delta _{j\alpha }+U^{jk}\delta
_{i\alpha }=-V^{k\alpha }\overleftarrow{M}_{ij},  \label{3.3} \\
&&U^{ij}\overleftarrow{Q}_{\alpha \beta }-W^{\alpha \beta }\overleftarrow{M}
_{ij}=V^{i\alpha }\delta _{j\beta }+V^{j\alpha }\delta _{i\beta }-V^{i\beta
}\delta _{j\alpha }-V^{j\beta }\delta _{i\alpha },  \label{3.4} \\
&&V^{i\alpha }\overleftarrow{P}_{j\beta }-V^{j\beta }\overleftarrow{P}
_{i\alpha }+V^{j\alpha }\delta _{i\beta }-V^{i\beta }\delta _{j\alpha }=0,
\label{3.5} \\
&&W^{\alpha \beta }\overleftarrow{P}_{i\gamma }-W^{\alpha \gamma }\delta
_{i\beta }+W^{\beta \gamma }\delta _{i\alpha }=V^{i\gamma }\overleftarrow{Q}
_{\alpha \beta },  \label{3.6} \\
&&W^{\alpha \beta }\overleftarrow{Q}_{\gamma \delta }+W^{\gamma \delta }
\overleftarrow{Q}_{\alpha \beta }=0,  \label{3.7}
\end{eqnarray}
where we used a notation
\begin{equation*}
U^{ij}\equiv F^{ij},\;V^{i\alpha }\equiv F^{i,n+\alpha },\;W^{\alpha \beta
}\equiv F^{n+\alpha ,n+\beta }.
\end{equation*}

I. It is obviously that Eq. (\ref{3.1_}) has solutions of the form
\begin{equation}
F^{AB}(p)=F_{1}^{AB}(p)=f(p)\overleftarrow{L}_{p}^{AB}  \label{F}
\end{equation}
where $f$ is an arbitrary polynomial in $p$.

II. The second solution has the form

\begin{eqnarray}
F^{AB}(p)=F_{2}^{AB}(p)&=&
(v_1+v_2\left\langle p,p\right\rangle)(-1)^{\varepsilon_A}\omega^{AB}=\nn
&=&(U_{2}^{ij}=0,\;V_{2}^{i\alpha
}(p)=(v_{1}+v_{2}\left\langle p,p\right\rangle )\delta _{i\alpha
},\;W_{2}^{\alpha \beta }=0),
\label{F2}
\end{eqnarray}
where $v_i$ are constant.
Note that $F_{2}^{AB}(p)$ can not be represented in the form (\ref{F}).

\lemma \label{ar.1}

General solution of Eq. (\ref{3.1_}) is
\begin{equation}
F^{AB}(p)=F_{1}^{AB}(p)+F_{2}^{AB}(p),  \label{3.9}
\end{equation}
where $F_1(p)$ and $F_2(p)$ are defined by Eqs. (\ref{F}) and (\ref{F2}).

\newpage
\appen{The proof of Proposition \ref{3.1_} for $n=1$.}

In this Appendix we omit everywhere the only possible index $1$.
Because $n=1$, Eqs. (\ref{3.2_}) - (\ref{3.7}) reduces to
\begin{eqnarray}
W=0\nn
\xi\partial_x U=0\nn
x\partial_x U -\xi\partial_\xi U +2U=-2\xi \partial_x V.\label{xxx}
\end{eqnarray}
So $U=u_1(x)\xi+u_0$. Let $V=\xi v_1(x)+v_0(x)$.
Then Eq. (\ref{xxx}) gives $u_0=0$ and
$x\partial_x u_1 +u_1=-2 \partial_x v_0$, which implies
$v_0(x)=-\frac 1 2 x u_1 + c$.

Let
$$f= -\frac 1 2 \int_0^x u_1\,dx+
\xi (x\partial_x -1)^{-1}[ v_1(x)-x\partial_y v_1(y)|_{y=0}].$$
Then $U=-2\xi\partial_x f$, $V=(x\partial_x -\xi\partial_\xi) f +c+
2c_1 \xi x$, where $c_1=-\frac 1 2 \partial_y v_1(y)|_{y=0}$.

\vskip 3mm

\appen{The proof of Proposition \ref{3.1_} for $n=2$.}

\vskip -3mm
\subparagraph{Equation (\protect\ref{3.7}):}

As $W$ is antisymmetric, the only nonzero element is $W^{12}=-W^{21}\equiv
w. $
It follows from (\ref{3.7}) that
$wQ^{12}=0$
Let $w=w_{12}\xi _{1}\xi _{2}+w_{1}\xi _{1}+w_{2}\xi _{2}+w_{0}$.
Then
$w\left( \overleftarrow{\frac{\partial }{\partial \xi _{1}}}x_{2}-
\overleftarrow{\frac{\partial }{\partial \xi _{2}}}x_{1}\right)
=-w_{12}x_{2}\xi _{2}-w_{12}x_{1}\xi _{1}+w_{1}x_{2}-w_{2}x_{1}=0$.
So
$w_{12}=0$
and
$w_{1}x_{2}-w_{2}x_{1}=0,$ which implies
$w_{i}=x_{i}W(x)$.
Thus
$W^{12}=W(x)(x_{1}\xi _{1}+x_{2}\xi _{2})+w_{0}(x)=(W(x)\xi _{2}\xi
_{1})Q^{12}+w_{0}(x)$.

Up to (\ref{F}) we have
\begin{equation*}
W^{12}(x,\xi )=w(x)
\end{equation*}

\subparagraph{Equation (\protect\ref{3.6}):}

As $W^{\alpha \beta }$ does not depend on $\xi $, we have
$x_{\gamma }\frac{\partial }{\partial x_{i}}W^{\alpha \beta }-W^{\alpha
\gamma }\delta _{i\beta }+W^{\beta \gamma }\delta _{i\alpha }=V^{i\gamma }
\overleftarrow{Q}_{\alpha \beta }$
.
Consider all 4 cases

$\alpha =1$, $\beta =2$, $i=1$, $\gamma =1$:
\begin{equation}
x_{1}\frac{\partial }{\partial x_{1}}w-w=V^{11}\overleftarrow{Q}^{12}
\label{11}
\end{equation}

$\alpha =1$, $\beta =2$, $i=2$, $\gamma =2$:
\begin{equation}
x_{2}\frac{\partial }{\partial x_{2}}w-w=V^{22}\overleftarrow{Q}^{12}
\label{22}
\end{equation}

$\alpha =1$, $\beta =2$, $i=2$, $\gamma =1$:
\begin{equation}
x_{2}\frac{\partial }{\partial x_{1}}w=V^{12}\overleftarrow{Q}^{12}
\label{21}
\end{equation}

$\alpha =1$, $\beta =2$, $i=1$, $\gamma =2$:
\begin{equation}
x_{1}\frac{\partial }{\partial x_{2}}w=V^{21}\overleftarrow{Q}^{12}
\label{12}
\end{equation}

All other equations from (\ref{3.6}) are equivalent to these ones.

The sum of the equations (\ref{11}) and (\ref{22}) gives $
(N_{x}-2)w=(V^{11}+V^{22})\overleftarrow{Q}^{12}$
and up to (\ref{F})
\begin{equation}
w(x)=\alpha _{11}x_{1}^{2}+\alpha _{12}x_{1}x_{2}+\alpha _{22}x_{2}^{2}
\label{w}
\end{equation}
where $\alpha _{ij}$ are constants.

Let us note that $\alpha _{12}x_{1}x_{2}=1/2\alpha _{12}\left( x_{1}\xi
_{1}-x_{2}\xi _{2}\right) \overleftarrow{Q}^{12}$ and we can regard that $
\alpha _{12}=0$ up to (\ref{F}).

Substitute (\ref{w}) to (\ref{11}), (\ref{22}), (\ref{21}) and (\ref{12})
and obtain
\begin{eqnarray*}
\alpha _{11}x_{1}^{2}-\alpha _{22}x_{2}^{2} &=&V^{11}\overleftarrow{Q}^{12}
\\
-\alpha _{11}x_{1}^{2}+\alpha _{22}x_{2}^{2} &=&V^{22}\overleftarrow{Q}^{12}
\\
2a_{11}x_{1}x_{2} &=&V^{12}\overleftarrow{Q}^{12} \\
2a_{22}x_{1}x_{2} &=&V^{21}\overleftarrow{Q}^{12}
\end{eqnarray*}
It follows from these equations that $V^{ij}$ does not contain the terms
proportional to $\xi _{1}\xi _{2}.$
So we can present $V^{ij}$ in the form
\begin{equation}
V^{ij}=v_{ij}^{1}(x)\xi _{1}+v_{ij}^{2}(x)\xi _{2}+v_{ij}^{0}(x)  \label{v}
\end{equation}
and obtain
\begin{eqnarray}
\alpha _{11}x_{1}^{2}-\alpha _{22}x_{2}^{2} &=&V^{11}\overleftarrow{Q}
^{12}=v_{11}^{1}(x)x_{2}-v_{11}^{2}(x)x_{1}  \label{v-11} \\
-\alpha _{11}x_{1}^{2}+\alpha _{22}x_{2}^{2} &=&V^{22}\overleftarrow{Q}
^{12}=v_{22}^{1}(x)x_{2}-v_{22}^{2}(x)x_{1}  \label{v-22} \\
2a_{11}x_{1}x_{2} &=&V^{12}\overleftarrow{Q}
^{12}=v_{12}^{1}(x)x_{2}-v_{12}^{2}(x)x_{1}  \label{v-12} \\
2a_{22}x_{1}x_{2} &=&V^{21}\overleftarrow{Q}
^{12}=v_{21}^{1}(x)x_{2}-v_{21}^{2}(x)x_{1}  \label{v-21}
\end{eqnarray}

These equations have the following partial solution for $v_{ij}^{\alpha }$:
\begin{eqnarray*}
v_{11}^{1}(x) &=&-\alpha _{22}x_{2}\ \ ;\ \ v_{11}^{2}(x)=-\alpha _{11}x_{1}
\\
v_{22}^{1}(x) &=&\alpha _{22}x_{2}\ \ ;\ \ v_{22}^{2}(x)=\alpha _{11}x_{1} \\
v_{12}^{1}(x) &=&0\ \ ;\ \ v_{12}^{2}(x)=-2\alpha _{11}x_{2} \\
v_{21}^{1}(x) &=&2\alpha _{22}x_{1}\ \ ;\ \ v_{21}^{2}(x)=0
\end{eqnarray*}
and the following general solution for $V^{ij}\overleftarrow{Q}^{12}=0$:
$v_{ij}^{\alpha }=x_{\alpha }v_{ij}(x)$.

So the general solution for $V^{ij}$ and $W^{12}$ obtained from (\ref{3.7})
and (\ref{3.6}) is
\begin{equation}
w(x)=\alpha _{11}x_{1}^{2}+\alpha _{22}x_{2}^{2}  \label{w sol}
\end{equation}
\begin{eqnarray}
V^{11} &=&f_{11}(x) \langle p,p \rangle +g_{11}(x)+(-\alpha _{11}x_{1}\xi _{2}-\alpha
_{22}x_{2}\xi _{1})  \label{v sol_} \\
V^{22} &=&f_{22}(x) \langle p,p \rangle +g_{22}(x)+(\alpha _{11}x_{1}\xi _{2}+\alpha
_{22}x_{2}\xi _{1})  \notag \\
V^{12} &=&f_{12}(x) \langle p,p \rangle +g_{12}(x)-2\alpha _{11}x_{2}\xi _{2}  \notag \\
V^{21} &=&f_{21}(x) \langle p,p \rangle +g_{21}(x)+2\alpha _{22}x_{1}\xi _{1}  \notag
\end{eqnarray}

\subparagraph{Equation (\protect\ref{3.5}):}

Let us substitute (\ref{v sol_}) to (\ref{3.5}):
$V^{i\alpha }\overleftarrow{P}_{j\beta }-V^{j\beta }\overleftarrow{P}
_{i\alpha }+V^{j\alpha }\delta _{i\beta }-V^{i\beta }\delta _{j\alpha }=0$
.
Consider the case

$i=1$, \ $\alpha =2$, \ $j=2$, \ $\beta =1$:
\begin{equation*}
V^{12}\overleftarrow{P}_{21}-V^{21}\overleftarrow{P}_{12}+V^{22}-V^{11}=0
\end{equation*}

$x_{1}\frac{\partial }{\partial x_{2}}f_{12}(x) \langle p,p \rangle +x_{1}\frac{\partial }{
\partial x_{2}}g_{12}(x)-2\alpha _{11}x_{1}\xi _{2}-(-2\alpha _{11}x_{2}\xi
_{1})-$

$-x_{2}\frac{\partial }{\partial x_{1}}f_{21}(x) \langle p,p \rangle -x_{2}\frac{\partial }{
\partial x_{1}}g_{21}(x)-2\alpha _{22}x_{2}\xi _{1}+(2\alpha _{22}x_{1}\xi
_{2})+$

$f_{22}(x) \langle p,p \rangle +g_{22}(x)+(\alpha _{11}x_{1}\xi _{2}+\alpha _{22}x_{2}\xi
_{1})-$

$-f_{11}(x) \langle p,p \rangle -g_{11}(x)+(\alpha _{11}x_{1}\xi _{2}+\alpha _{22}x_{2}\xi
_{1})=0$.

Extract from the left hand side polynomial the terms $x_{2}\xi _{1}$ and $
x_{1}\xi _{2}$ :

$2\alpha _{11}x_{2}\xi _{1}=0$

$2\alpha _{22}x_{1}\xi _{2}=0$

So
\begin{equation*}
\alpha _{11}=\alpha _{22}=0
\end{equation*}

\subparagraph{Equation (\protect\ref{3.5}) again:}

Now $W^{12}=0,$ $V^{ij}=f_{ij}(x) \langle p,p \rangle +g_{ij}(x).$

There exist the solutions of the equations
\begin{eqnarray}
V^{ij} &=&\left( f(x) \langle p,p \rangle +g(x)\right) \overleftarrow{P}_{ij}=x_{j}\frac{
\partial f(x)}{\partial x_{i}} \langle p,p \rangle +x_{j}\frac{\partial g(x)}{\partial x_{i}}
\label{fg} \\
W^{12} &=&0  \notag \\
U^{ij} &=&-\left( f(x) \langle p,p \rangle +g(x)\right) \overleftarrow{M}_{ij}  \notag
\end{eqnarray}

Let us look for the solution of (\ref{3.5}) up to (\ref{fg}).
As the function $ \langle p,p \rangle $ commutes with all the operators $L^{AB},$ one can
consider (\ref{3.5}) for $f_{ij}$ and $g_{ij}$ separately, namely
\begin{equation*}
x_{l}\partial _{k}f_{ij}-x_{j}\partial _{i}f_{kl}+f_{kj}\delta
_{il}-f_{il}\delta _{kj}=0
\end{equation*}
Consider 3 cases:

1122)

$i=1$, \ $j=1$, \ $k=2$, \ $l=2$:

\begin{equation*}
x_{2}\partial _{2}f_{11}-x_{1}\partial _{1}f_{22}=0
\end{equation*}
So $\partial _{2}f_{11}=x_{1}\varphi (x)$, \ $\partial
_{1}f_{22}=x_{2}\varphi (x)$ and

$f_{11}=\int_0^{x_2} x_{1}\varphi (x)dx_{2}+\varphi _{1}(x_{1})$,

$f_{22}=\int_0^{x_1} x_{2}\varphi (x)dx_{1}+\varphi _{2}(x_{2})$,

So

$f_{11}=x_{1}\frac{\partial }{\partial x_{1}}\left(
\int_0^{x_1} \int_0^{x_2} \varphi
(x)dx_{1}dx_{2}+N_{x}^{-1}\varphi _{1}(x_{1})+N_{x}^{-1}\varphi
_{2}(x_{2})\right) +c_{11}$,

$f_{22}=x_{2}\frac{\partial }{\partial x_{2}}\left(
\int_0^{x_1} \int_0^{x_2} \varphi
(x)dx_{1}dx_{2}+N_{x}^{-1}\varphi _{1}(x_{1})+N_{x}^{-1}\varphi
_{2}(x_{2})\right) +c_{22}$,

where $c_{ii}$ are constant.

1112) and 2212)

$i=1$, \ $j=1$, \ $k=1$, \ $l=2$:

\begin{equation*}
x_{2}\partial _{1}f_{11}-x_{1}\partial _{1}f_{12}-f_{12}=0
\end{equation*}

$i=2$, \ $j=2$, \ $k=1$, \ $l=2$

\begin{equation*}
x_{2}\partial _{1}f_{22}-x_{2}\partial _{2}f_{12}+f_{12}=0
\end{equation*}
The sum of these equations gives

$N_{x}f_{12}=x_{2}\partial _{1}(f_{11}+f_{22})=x_{2}\partial _{1} N_{x}\left(
\int_0^{x_1} \int_0^{x_2} \varphi (x)dx_{1}dx_{2}+N_{x}^{-1}\varphi
_{1}(x_{1})+N_{x}^{-1}\varphi _{2}(x_{2})\right) $.

So
$f_{12}=x_{2}\partial _{1}\left( \int_0^{x_1} \int_0^{x_2} \varphi
(x)dx_{1}dx_{2}+N_{x}^{-1}\varphi _{1}(x_{1})+N_{x}^{-1}\varphi
_{2}(x_{2})\right) +c_{12}$, where $c_{12}$ is constant.

Thus we obtain that up to (\ref{fg}) the solution of (\ref{3.5}) has the form
\begin{equation*}
V^{ij}=a_{ij} \langle p,p \rangle +b_{ij}
\end{equation*}
with constant $a_{ij}$, $b_{ij}. $Substituting this expression to (\ref{3.5})
we finally obtain
\begin{equation*}
V^{ij}=a\delta _{ij} \langle p,p \rangle +b\delta _{ij}
\end{equation*}

\subparagraph{Equations (\protect\ref{3.2_}), (\protect\ref{3.3}) and (
\protect\ref{3.4}):}

When $W^{ij}=0$ and $V^{ij}=a\delta _{ij} \langle p,p \rangle +b\delta _{ij}$, the equations
(\ref{3.2_}), (\ref{3.3}) and (\ref{3.4}) take the form
\begin{eqnarray}
&&U^{ij}\overleftarrow{M}_{kl}+U^{kl}\overleftarrow{M}_{ij}=0,  \label{u1} \\
&&U^{ij}\overleftarrow{P}_{k\alpha }+U^{ik}\delta _{j\alpha }+U^{jk}\delta
_{i\alpha }=0  \label{u2} \\
&&U^{ij}\overleftarrow{Q}_{\alpha \beta }=0  \label{u3}
\end{eqnarray}

The solution of (\ref{u3}) has the form
\begin{equation}
U^{ij}=u_{ij}^{1}(x) \langle p,p \rangle +u_{ij}^{0}(x)  \label{u3s}
\end{equation}
and the equations (\ref{u1}) and (\ref{u2}) can be considered for $u^{\alpha
} $ separately.
Eq. (\ref{u2}) gives

\begin{eqnarray*}
u_{11}^{\alpha }\overleftarrow{P}_{22} &=&0 \\
u_{11}^{\alpha }\overleftarrow{P}_{11}+2u_{11}^{\alpha } &=&0
\end{eqnarray*}
which implies $(N_{x}+2)u_{11}^{\alpha }=0$ and as a consequence $
u_{11}^{\alpha }=0$ (because $u_{11}^{\alpha }$ is a polynomial).
Analogously $u_{22}^{\alpha }=0.$

Further, Eq. (\ref{u2}) gives
\begin{eqnarray*}
u_{12}^{\alpha }\overleftarrow{P}_{11}+u_{21}^{\alpha } &=&0 \\
u_{12}^{\alpha }\overleftarrow{P}_{22}+u_{12}^{\alpha } &=&0
\end{eqnarray*}
Taking in account the relation $u_{12}^{\alpha }=u_{21}^{\alpha },$ the sum
of the last 2 equations gives
$(N_{x}+2)u_{12}^{\alpha }=0$ and $u_{12}^{\alpha }=0.$

So, up to (\ref{F}), the solution of the equations (\ref{3.2_})-(\ref{3.7})
has the form
\begin{eqnarray*}
U^{ij} &=&0 \\
V^{ij} &=&a\delta _{ij} \langle p,p \rangle +b\delta _{ij}\text{ with constant }a\text{ and }
b \\
W^{ij} &=&0
\end{eqnarray*}

\appen{The proof of Proposition \ref{3.1_} for $n>2$.}

The case $n>2$ we prove using induction hypothesis.

Let for $n\leq N-1$ lemma is true.
Consider the case $n=N.$
According the inductive hypothesis we can regard, that up (\ref{F}) the
solution
has the form
\begin{eqnarray}
W^{ij} &=&0\ \ \text{for }i,j=2,...,N  \label{Sol N-1} \\
V^{ij} &=&\delta _{ij}(v_{0}(x_{1},\xi _{1})+v_{1}(x_{1},\xi _{1}) \langle p,p \rangle )\ \
\text{for }i,j=2,...,N  \notag \\
U^{ij} &=&0\ \ \text{for }i,j=2,...,N  \notag
\end{eqnarray}

\subparagraph{Equation (\protect\ref{3.7}):}

The equation (\ref{3.7}) gives

\begin{equation}
W^{1i}\overleftarrow{Q}_{jk}=W^{1i}\left( \frac{\overleftarrow{\partial }}{
\partial \xi _{j}}x_{k}-\frac{\overleftarrow{\partial }}{\partial \xi _{k}}
x_{j}\right) =0\ \ \text{for }i,j,k=2,...,N  \label{Wij}
\end{equation}

It is evident that $W^{1i}$ is at most linear on $\xi _{s}$ for all $\xi
_{s} $ with $s>1$\ due to (\ref{Wij}). Decompose $W^{1i}$:
\begin{equation*}
W^{1i}=w_{0}^{i}(x,\xi _{1})+\sum_{s=2}^{K}w_{s}^{i}(x,\xi _{1})\xi _{s}
\end{equation*}

Eq. (\ref{3.7}) gives
\begin{equation*}
w_{j}^{i}(x,\xi _{1})x_{k}=w_{k}^{i}(x,\xi _{1})x_{j},
\end{equation*}
which implies
\begin{equation*}
w_{j}^{i}(x,\xi _{1})=2w^{i}(x,\xi _{1})x_{j},
\end{equation*}
which implies in its turn
\begin{equation*}
W^{1i}=W_{0}^{i}(x,\xi _{1})+W_{1}^{i}(x,\xi _{1}) \langle p,p \rangle .
\end{equation*}

Consider the remaining equations from Eq. (\ref{3.7}):
\begin{equation*}
W^{1i}\overleftarrow{Q}_{1k}=\left( W_{0}^{i}(x,\xi _{1})+W_{1}^{i}(x,\xi
_{1}) \langle p,p \rangle \right) \left( \frac{\overleftarrow{\partial }}{\partial \xi _{1}}
x_{k}-\frac{\overleftarrow{\partial }}{\partial \xi _{k}}x_{1}\right) =0\ \
\text{for }i,k=2,...,N
\end{equation*}
which implies that $W_{\alpha }^{i}$ do not depend on $\xi _{1}$.

So
\begin{equation}
W^{1i}=\left( W_{0}^{i}(x)+W_{1}^{i}(x) \langle p,p \rangle \right) \ \ \text{for }i=2,...,N
\label{WN}
\end{equation}

\subparagraph{Equation (\protect\ref{3.6}):}

The equation (\ref{3.6}) gives
\begin{equation}
W^{\alpha \beta }\overleftarrow{P}_{i\gamma }-W^{\alpha \gamma }\delta
_{i\beta }+W^{\beta \gamma }\delta _{i\alpha }=V^{i\gamma }\overleftarrow{Q}
_{\alpha \beta }  \label{3.6.1}
\end{equation}
where $\overleftarrow{Q}_{1\beta }=\frac{\overleftarrow{\partial }}{\partial
\xi _{1}}x_{\beta }-\frac{\overleftarrow{\partial }}{\partial \xi _{\beta }}
x_{1}$ ($\beta >1$).

The case $\alpha >1,$ $\beta >1,$ $i=1,\gamma \geq 1$ gives
\begin{equation}
V^{1i}=\left( V_{0}^{i}(x,\xi _{1})+V_{1}^{i}(x,\xi _{1}) \langle p,p \rangle \right) \ \
\text{for }i=1,2,...,N  \label{VN}
\end{equation}

The case $\alpha =1,$ $\beta >1,$ $i=\gamma $ gives
\begin{equation*}
\left( N_{x}-2\right) W^{1\beta }=\left( \sum_{i=1}^{N}V^{ii}\right)
\overleftarrow{Q}_{1\beta }
\end{equation*}

It follows from (\ref{VN}) that $\left( \sum_{i=1}^{N}V^{ii}\right)
\overleftarrow{Q}_{\alpha \beta }=0$ for $\alpha ,\beta >1$. So, up to (\ref
{F}) we have that $W_{0}^{i}(x)$ and $W_{1}^{i}(x)$ are polynomials of the
second order.
\begin{equation}
W^{1i}=\left( W_{0}^{i}(x)+W_{1}^{i}(x) \langle p,p \rangle \right) \ \ \text{for }i=2,...,N
\label{W2}
\end{equation}

Evidently, $W_{0}^{i}(x)$ and$\ W_{1}^{i}(x)$ satisfy (\ref{3.6}) separately.

The case $\alpha =1,$ $\beta >1,$ $i>1,\gamma =\beta \neq i$ gives
\begin{equation*}
W_{k}^{\beta }(x)\frac{\overleftarrow{\partial }}{\partial x_{i}}x_{\beta }=0
\end{equation*}
and so $W_{k}^{\beta }$ depends on $x_{1}$ and $x_{\beta }$ only.

The case $\alpha =1,$ $\beta =i>1,$ $i\neq \gamma >1$ gives $W_{k}^{\beta
}(x)\frac{\overleftarrow{\partial }}{\partial x_{\beta }}x_{\gamma
}-W_{k}^{\gamma }=0.$

So
\begin{equation*}
W_{k}^{\beta }=c_{k}x_{1}x_{\beta }
\end{equation*}
and $W^{\alpha \beta }=(c_{0}x_{1}\xi _{1}+c_{1}x_{1}\xi _{1} \langle p,p \rangle )
\overleftarrow{Q}_{\alpha \beta }.$ Because $(c_{0}x_{1}\xi
_{1}+c_{1}x_{1}\xi _{1} \langle p,p \rangle )\overleftarrow{L}^{AB}$ has the form (\ref{Sol
N-1}) we can regard that \ $W^{\alpha \beta }=0$ up to (\ref{F}).

With $W^{\alpha \beta }=0$ the equation (\ref{3.6.1}) gives $V^{ij}
\overleftarrow{Q}_{1\beta }=0$ which implies that
\begin{equation}
V^{ij}=\left( V_{0}^{ij}(x)+V_{1}^{ij}(x) \langle p,p \rangle \right) \ \ \text{for }
i,j=1,2,...,N
\end{equation}
where $V_{k}^{ij}=0$ if $1<i\neq j>1.$

So the equation (\ref{3.5}) is valid for $V_{0}^{ij}(x)$ and $V_{1}^{ij}(x)$
separately.

\subparagraph{Equation (\protect\ref{3.5}):}

The equation (\ref{3.5}) has the following form for the functions depending
on $x$ only:
\begin{equation*}
x_{\beta }\frac{\partial }{\partial x_{j}}v^{i\alpha }-x_{\alpha }\frac{
\partial }{\partial x_{i}}v^{j\beta }+v^{j\alpha }\delta _{i\beta
}-v^{i\beta }\delta _{j\alpha }=0
\end{equation*}

Consider the case $i=\alpha =1$,  $j=$ $\beta >1$. We have $x_{j}\frac{
\partial }{\partial x_{j}}v^{11}-x_{1}\frac{\partial }{\partial x_{1}}
v^{jj}=0$. Because $v^{jj}(x)$ does not depend on $x_{j}$ for $j>1$ we have $
v^{jj}=cons\dot{t}$ and $\frac{\partial }{\partial x_{j}}v^{11}=0.$

Consider the case $i=1$, $j>1$, $\beta >1$, $\beta \neq j=\alpha .$We have $
x_{\beta }\frac{\partial }{\partial x_{j}}v^{1j}-v^{1\beta }=0.$

Consider the case $\alpha =1$, $\ i=j=\beta >1.$We have $x_{i}\frac{\partial
}{\partial x_{i}}v^{i1}+v^{i1}=0$ which implies $v^{i1}=0$ because $v^{i1}$
are polynomials.

Consider the case $i=1$, $\alpha =$ $j=$ $\beta >1$. We have $x_{j}(\frac{
\partial }{\partial x_{j}}v^{1j}-\frac{\partial }{\partial x_{1}}
v^{jj})+v^{jj}=0.$ Because $v^{jj}=const$ this gives $v^{jj}=0.$ So $\frac{
\partial }{\partial x_{j}}v^{1j}=0.$

Consider the case $i=\alpha =j=1,$ $\beta >1.$ We have $x_{\beta }\frac{
\partial }{\partial x_{1}}v^{11}-x_{1}\frac{\partial }{\partial x_{1}}
v^{1\beta }-v^{1\beta }=0$ which gives $v^{1\beta }=x_{\beta }v(x_{1})$ and
finally $v^{1\beta }=0,$ $v^{11}=const.$

Thus, $V^{ij}=(c_{0}+c_{1} \langle p,p \rangle )\delta _{ij}.$

\subparagraph{Equations (\protect\ref{3.2_}), (\protect\ref{3.3}) and (\ref{3.4}):}

\begin{eqnarray}
&&U^{ij}\overleftarrow{M}_{kl}+U^{kl}\overleftarrow{M}_{ij}=0,  \label{3.2.0}
\\
&&U^{ij}\overleftarrow{P}_{k\alpha }+U^{ik}\delta _{j\alpha }+U^{jk}\delta
_{i\alpha }=0,  \label{3.3.0} \\
&&U^{ij}\overleftarrow{Q}_{\alpha \beta }=0  \label{3.4.0}
\end{eqnarray}
where $U^{ij}=0\ \ $for $i,j=2,...,N.$

Consider the equation (\ref{3.3.0}) at $k=1,$ $i=j=\alpha >1.$ We have $
2U^{1i}=0$.

Consider the equation (\ref{3.3.0}) at $i=k=1,$ $j=\alpha >1.$ We have $
U^{11}=0$.

Thus, up to (\ref{F})
\begin{eqnarray*}
W^{ij} &=&0 \\
V^{ij} &=&(c_{0}+c_{1} \langle p,p \rangle )\delta _{ij} \\
U^{ij} &=&0
\end{eqnarray*}

\end{document}